\documentclass[%
 aip,
 amsmath,amssymb,
 reprint,longbibliography,
]{revtex4-1}
\usepackage[utf8]{inputenc}
\usepackage{graphicx}
\usepackage{bm}
\usepackage{bbold}
\usepackage{cases}
\usepackage{upgreek}
\usepackage[usenames, dvipsnames]{color}
\usepackage[colorlinks=true, citecolor=Plum, linkcolor=BurntOrange, urlcolor=Plum ]{hyperref}
\usepackage{braket}
\usepackage[normalem]{ulem}
\usepackage{xfrac}

\usepackage{textgreek}
\usepackage{gensymb}
\usepackage{dcolumn}

\newcommand \mc[1] { \mathcal{#1} }
\newcommand \dd[1]  { \!\!\textrm d{#1} \,}   
\newcommand \rmm[1]  { \textrm{#1} }

\newcommand \e[1] { \rmm{e}^{#1} }

\newcommand \ris { \!\!\! }

\makeatletter
\def\@email#1#2{%
 \endgroup
 \patchcmd{\titleblock@produce}
  {\frontmatter@RRAPformat}
  {\frontmatter@RRAPformat{\produce@RRAP{*#1\href{mailto:#2}{#2}}}\frontmatter@RRAPformat}
  {}{}
}%
\makeatother
\begin{document}

\preprint{AIP/123-QED}

\title{Efficient formulation of multitime generalized quantum master equations: \\ Taming the cost of simulating 2D spectra}

\author{Thomas Sayer}
\affiliation{Department of Chemistry, University of Colorado Boulder, Boulder, CO 80309, USA\looseness=-1}

\author{Andr\'{e}s Montoya-Castillo}
\homepage{Andres.MontoyaCastillo@colorado.edu}
\affiliation{Department of Chemistry, University of Colorado Boulder, Boulder, CO 80309, USA\looseness=-1} 


\date{\today}

\begin{abstract}

Modern 4-wave mixing spectroscopies are expensive to obtain experimentally and computationally. In certain cases, the unfavorable scaling of quantum dynamics problems can be improved using a generalized quantum master equation (GQME) approach. However, the inclusion of multiple (light-matter) interactions complicates the equation of motion and leads to seemingly unavoidable cubic scaling in time. In this paper, we present a formulation that greatly simplifies and reduces the computational cost of previous work that extended the GQME framework to treat arbitrary numbers of quantum measurements. Specifically, we remove the time derivatives of quantum correlation functions from the modified Mori-Nakajima-Zwanzig framework by switching to a discrete-convolution implementation inspired by the transfer-tensor approach. We then demonstrate the method's capabilities by simulating 2D electronic spectra for the excitation-energy-transfer dimer model. In our method, the resolution of the data can be arbitrarily coarsened, especially along the $t_2$ axis, which mirrors how the data are obtained experimentally. Even in a modest case, this demands $\mathcal{O}(10^3)$ fewer data points. We are further able to decompose the spectra into 1-,~2-, and 3-time correlations, showing how and when the system enters a Markovian regime where further measurements are unnecessary to predict future spectra and the scaling becomes quadratic. This offers the ability to generate long-time spectra using only short-time data, enabling access to timescales previously beyond the reach of standard methodologies.

\end{abstract}

\maketitle

\section{Introduction}
\vspace{-6pt}

Multidimensional spectroscopies are powerful tools in the natural sciences that employ multiple light pulses in particular geometries, phases, and frequencies to reveal critical information about the structure and function of complex materials.\cite{Mukamel1999a, Jonas2003a, Cho2009, Cho2008a, Schlau-Cohen2011a, Jonas2016, Biswas2022a} Specifically, the use of several photons can generate excited states and exotic quasiparticles, while control over their relative timings, interferences, and energies reveals the underlying dynamics, including pathways for chemical reactivity, energy relaxation, and transfer. The ultimate goal of theory is therefore to meet the challenge of this complex yet insightful methodology with an equally detailed and accurate quantum dynamical description. Yet, computing spectral responses in the form of multitime correlation functions over the relevant timescales probed by experiments is prohibitively expensive with all but the most approximate methods or the simplest models and regimes. Thus, developing a generally applicable and affordable method to calculate long-time correlations in quantum systems subject to measurements represents a formidable and important challenge in quantum dynamics.

Modern multidimensional spectroscopic methods use a 4-wave mixing technique whose three independent waiting times demand a workflow with cubic scaling in time. The goal is then to acquire dynamical knowledge of some set of correlation functions over their full range. For example, 2D electronic spectra probe processes ranging from ultrafast electron transfer on the order of femtoseconds, to hole trapping on nanocrystal surfaces over many picoseconds, to conformational changes in photoprotection mechanisms in plants occurring over nanoseconds.\cite{Engel2007b, Cho2010, Lewis2012, Rafiq2015, Jonas2016, Ryu2021, Son2021, Son2021a, Policht2022, Biswas2022a, Nguyen2023, Brosseau2023} The length of these correlation times, when compared to the much finer timesteps required to generate their underlying dynamics, make predicting spectral responses a significant computational challenge. This problem is exacerbated by the fact that most numerically exact and even approximate methods scale nonlinearly, and at worst exponentially, with simulation time.\cite{Chen2010h, Liang2014a, Yan2021c, Jansen2021a, Chen2022b, Segatta2023} Here we suggest that in such cases---in analogy to modern approaches to simulating 1-time correlation functions and nonequilibrium averages\cite{Shi2003a, Shi2004b, Zhang2006, Cohen2011a, Montoya-Castillo2016, Kelly2016, Kidon2018a, Mulvihill2021a}---one can drastically reduce the computational cost of simulating multitime correlation functions by employing a generalized quantum master equation (GQME). 

The Mori-Nakajima-Zwanzig (MNZ) equation is a `1-time GQME' that describes the evolution of a low-dimensional set of observables, $\bm{\mc{C}}(t)$, by formally projecting them out of the generally high- or even infinite-dimensional Hilbert space of the full system.\cite{Nakajima1958a, Zwanzig1960, Mori1965b} While the evolution of the entire system is Markovian (i.e., where the next step in the evolution only requires information from the present step), the cost of reducing the dimensionality is that evolving $\bm{\mc{C}}(t)$ requires knowledge of some or even all of its dynamical history. This non-Markovianity is contained in the last term of the MNZ equation,
\vspace{-6pt}
\begin{equation}\label{eq:NZ}
    \dot{\bm{\mc{C}}}(t) = \dot{\bm{\mc{C}}}(0) \bm{\mc{C}}(t) + \bm{\mc{I}}(t) - \int_0^t \dd{\tau}\, \bm{\mc{K}}(\tau) \bm{\mc{C}}(t-\tau),
\end{equation}
where the memory kernel, $\bm{\mc{K}}(\tau)$, appears under a convolution integral. The second term, $\bm{\mc{I}}(t)$, arises from the time-zero correlation of the observables of interest in the projection operator, $\mc{P}$, with the `unimportant' degrees of freedom that are to be integrated out. A commonly used assumption is that the initial condition of the entire system is multiplicatively separable, $\hat{\Gamma}(0) \approx \hat{\rho}_B \otimes \hat{\rho}_S$. This is appropriate for spectroscopies with impulsive excitations. When treating linear electronic spectroscopy, one can employ a projection operator that tracks the dynamics of the electronic states while integrating out all nuclear motions that modulate their dynamics. This ensures that the inhomogeneous term is zero, $\bm{\mc{I}}(t) = 0$. In such cases, the only important object is the memory kernel, $\bm{\mc{K}}(t)$. In dissipative systems, the memory kernel goes to zero after some finite time $\tau_K$, which places an upper limit on the amount of dynamical information required to fully characterize the projected system.

The complexity increases when a correlation function needs to include interactions at intermediate times, such as the additional light pulses $B_1, B_2,...,B_n$ in multidimensional spectroscopies. In effect, every time an interaction takes place during the experiment, a new `initial condition' is generated that subsequently propagates in time. For instance, one intermediate interaction at $t_1$ defines a new 2-time propagator, $\bm{\mc{C}}^{B_1}(t_2, t_1)$. For this new initial condition, the total system density is no longer multiplicatively separable, which ultimately leads the multitime GQME to be encumbered by the 2-time equivalent of the inhomogeneous term. Therefore, one cannot assume that the multitime correlation function can be composed using the solution to the 1-time problem: $\bm{\mc{C}}^{B_1}(t_2, t_1) \neq \bm{\mc{C}}(t_2)\bm{\mc{B}}_1\bm{\mc{C}}(t_1)$. Additional interactions continue to generate such complicating terms in ever higher dimensions, which makes applying the MNZ equation to multitime problems like 2D spectroscopy fundamentally more complex than for single-photon experiments. Nevertheless, it would be advantageous to extend and exploit the significant improvements in efficiency GQMEs can offer for 1-time correlation functions\cite{Shi2004b, Kelly2013a, Kelly2015a, Montoya-Castillo2016, Kelly2016, Mulvihill2019b, Mulvihill2019c, Amati2022a} for their multitime counterparts, which are essential in predicting nonlinear responses.

While important challenges remain, recent work has significantly advanced this perspective. Specifically, Ivanov and Breuer have developed a projection operator-based decomposition of the reduced propagator after $n$ measurements, $\bm{\mc{C}}^{B_n, B_{n-1},...,B_1}(t_{n+1}, t_n, ..., t_1)$, that encapsulates all measurement-induced inhomogeneities in multitime memory kernels.\cite{Ivanov2015a} This important advance offers the means to separate contributions of higher-order correlation functions (with multiple time indices) from those constructed as products of lower orders. Here, we show how such an approach enables one to quantify the extent to which the additional light pulses introduced in multidimensional setups contain novel information not accessible from simpler experiments. 

Yet, the continuous-time formulation of this approach retains large computational expenses that limit its applicability. First, for a 4-wave mixing experiment, the method requires the precalculation of the 3-time propagator, $\bm{\mc{C}}^{B_2,B_1}(t_3, t_2, t_1)$, as well as related 2-time quantities for sufficiently long times to enable the construction of the single- and multitime memory kernels. Crucially, the post-processing takes place in the quasi-continuous time domain, which requires high resolution to keep integral discretization errors small. In contrast, experimental spectra do not require such high resolution. Indeed, it is common to report only a few waiting times along $t_2$ in 2D spectroscopy. To that end, many theoretical studies only simulate spectra at specific $t_2$ values, which leads to a quadratic (rather than cubic) scaling in time.\cite{Chen2010h, Chen2011a, Hein2012b, Ikeda2017a, Leng2018a, Yan2021c, Chen2022b} In contrast, extension in $t_2$ using the quasi-continuous decomposition method requires the same high-resolution data along \textit{all} $t$-axes. Therefore, although the GQME method can reduce the total required simulation time along each axis, its use can in some cases \textit{increase} the cost relative to those previous studies by increasing the scaling in time. Second, in the original formulation,\cite{Ivanov2015a} memory kernels are obtained via the Dyson identity (see Eq.~\ref{eq:dyson_identity}). This introduces numerical derivatives into the expressions which limits their accuracy in the presence of noise or, as just discussed, in the regime of low temporal resolution.\cite{Sayer2023} Hence, despite the conceptual advantages offered by the quasi-continuous decomposition method, its computational cost can be even greater than a direct solution to the problem. 

To remove these limitations, we present a discrete-time reformulation of the method, with a view to performing 2D spectroscopy. To accomplish this, we leverage the mathematics of the transfer tensor method (TTM),\cite{Cerrillo2014a} which offers a discrete-time analog to the 1-time MNZ equation. In the original formulation with a vanishing inhomogeneous term, $\bm{\mc{I}}(t)=\bm{0}$, the discrete-time transfer tensor, $\bm{\mc{T}}_{n}$, has the same lifetime as the continuous-time memory kernel in the MNZ equation, $\bm{\mc{K}}(t)$. However, because the TTM directly decomposes the full dynamical object of interest rather than an MNZ-like equation obtained via projection operator techniques, it subsumes the effect of the $\bm{\mc{K}}(t)$ and $\bm{\mc{I}}(t)$ into the transfer tensor. The result is that the transfer tensor loses its time translational invariance when $\bm{\mc{I}}(t) \neq \bm{0}$, requiring two time indices rather than one, and has the lifetime of the longest-lived MNZ quantity, either $\bm{\mc{K}}(t)$ or $\bm{\mc{I}}(t)$. This impedes an unambiguous assignment of memory and inhomogeneous terms in the resulting TTM decomposition, thereby sacrificing the ability to separate the lifetime of correlations arising from the true initial conditions versus those from intermediate interactions in the case of multitime quantities. For this reason, we take a different approach from those adopted by recent work that generalizes the TTM procedure to directly decompose multitime correlation functions into multi-indexed transfer tensors.\cite{Jørgensen2020a, Gherardini2022}

Specifically, we exploit the mathematical power of the TTM-like decomposition of convolution integrals to reformulate the continuous-time projection operator-based decomposition of multitime correlation functions\cite{Ivanov2015a} to discrete counterparts with far lower computational cost and no loss of accuracy. Our development enables us to considerably reduce computational expense while preserving the clear interpretability of the projection operator-based decomposition in terms of 1-, 2-, and 3-time contributions to the total spectrum. 

The remainder of the paper introduces our new theoretical framework in three parts. We first provide an exposition of the relevant mathematics in Sec.~\ref{sec:section2}. Then, in Sec.~\ref{sec:spin-boson_results}, we describe the practical implementation of this discrete-time method using the spin-boson model as a validation tool. Finally, in Sec.~\ref{sec:eet_results}, we demonstrate the improvements in computational efficiency and novel physical insight that it can offer for 2D electronic spectroscopy using the excitation-energy-transfer (EET) dimer Hamiltonian as an illustrative example.

\vspace{-14pt}
\section{Theory}\label{sec:section2}
\vspace{-6pt}

Here we first provide a summary and analysis of the projection operator-based decomposition of multitime correlation functions presented in Ref.~\onlinecite{Ivanov2015a} in our notation and discuss its implications. This rewriting of the problem sheds light on how to calculate the 2D spectrum. We then separately introduce the TTM approach and then demonstrate how we employ convolutional discretizations to offer a highly efficient discrete-time representation of the projection operator-based decompositions of the 1-, 2- and 3-time reduced propagators.

\vspace{-8pt}
\subsection{Multitime Memory Kernel Theory}\label{sec:multi_theory}
\vspace{-6pt}

We begin by focusing on a 2-time correlation function where an interaction occurs, separating the evolution into two time ranges, $t_1$ and $t_2$. In our notation, this corresponds to a time-dependent correlation \textit{matrix} where a measurement $B$ is made after a time $t_1$:
\begin{equation}\label{eq:2timeprop}
    \boldsymbol{\mc{C}}^B(t_2, t_1) = (\mathbf{A}|\e{\mc{L}t_2} B \e{\mc{L}t_1}|\mathbf{A}),
\end{equation}
where the Liouvillian, $\mc{L}$, is kept general as the current framework applies to both quantum or classical systems. In the quantum case that we are considering, however, $\mc{L} \equiv -i[H, ...]$. Here, $(A_j|X|A_k) \equiv \mathrm{Tr}[A_jXA^{\dagger}_k\rho_B]$, $\rho_B$ is the density of nuclear modes, the operators $A_j \in \{ \ket{n}\bra{m}\}$ span the Liouville space of the electronic degrees of freedom, and $B$ is assumed to be an operator in the electronic space. The next step is to decompose the system operator, $B$, into two terms
\begin{equation}\label{eq:system_split}
    B = \mc{P} B \mc{P} + \mc{Q} B \mc{Q},
\end{equation}
where $\mc{P} = |\mathbf{A})(\mathbf{A}|$ and $\mc{Q} = \mathbb{1} - \mc{P}$. The cross terms ($\mc{Q} B \mc{P}$ and $\mc{P} B \mc{Q}$) in Eq.~\ref{eq:system_split} are equal to zero since $B$ can be expressed using the $\{A_j\}$ basis that composes $\mc{P}$. Substituting Eq.~\ref{eq:system_split} into Eq.~\ref{eq:2timeprop} and Dyson expanding the propagators (see Eq.~\ref{eq:dyson_identity}) in the $\mc{Q}B\mc{Q}$ term one obtains
\begin{equation}\label{eq:2timeprop_decomp}
\begin{split}
    \boldsymbol{\mc{C}}^B(t_2, t_1) &=  \boldsymbol{\mc{C}}(t_2) \mc{B} \boldsymbol{\mc{C}}(t_1) \\
    &~~+ \boldsymbol{\mc{C}}(t_2 - \tau_2) \boldsymbol{\mc{K}}^B(\tau_2, \tau_1) \boldsymbol{\mc{C}}(t_1 - \tau_1),
\end{split}
\end{equation}
where $\boldsymbol{\mc{B}} \equiv (\mathbf{A}|B|\mathbf{A})$ and we have employed a continuous version of the Einstein summation convention where a repeated $\tau_i$ index implies an integral with respect to $\tau_i$ from 0 to $t_i$. That is, with the convolution integrals written explicitly, the last term of Eq.~\ref{eq:2timeprop_decomp} is equivalent to 
\begin{equation}\label{eq:2timeprop_decomp-explicit}
\begin{split}
    \int_0^{t_1} d\tau_1 \int_0^{t_2} d\tau_2\ \boldsymbol{\mc{C}}(t_2 - \tau_2) \boldsymbol{\mc{K}}^B(\tau_2, \tau_1) \boldsymbol{\mc{C}}(t_1 - \tau_1). \nonumber
\end{split}
\end{equation}
The 2-time memory kernel for the measurement $B$ is
\begin{equation}\label{eq:2timemem}
    \bm{\mc{K}}^B(\tau_2, \tau_1) \equiv (\mathbf{A}|\mc{L} \e{\mc{QL}\tau_2}  \mc{Q} B \mc{Q} \e{\mc{LQ}\tau_1} \mc{L} |\mathbf{A}),
\end{equation}
which is difficult to evaluate owing to the presence of projected propagators, $\e{\mc{QL}t}$. 

Physically, the multitime memory kernel decomposition in Eq.~\ref{eq:2timeprop_decomp} can be understood as follows: The first term describes the dynamics if there were no system-bath correlation at the time of the measurement. We refer to this term as the `homogeneous' term, as it has no information regarding the complications arising from the measurements, meaning it be constructed using only 1-time data. Note that this term is written as time-local, but is not Markovian since the 1-time propagator is the object which satisfies the original MNZ equation. In fact, Eq.~\ref{eq:2timeprop_decomp} is not strictly a master equation. Although its multitime kernels are homologous to those that emerge from the projection operator approach to derive the projected equation of motion, Eq.~\ref{eq:2timeprop_decomp} more closely resembles a generalization to the Gaussian decomposition of multitime correlation functions. More specifically, it exactly unfurls the dynamical behavior of a multitime correlation function in terms of its Gaussian-like decomposition into products of 1-time terms and their multitime corrections.\cite{BookHansenMcdonald, BookBoonYip, Rahman1964, Kob1997, VanZon2001, Szamel2003, Wu2003b, Mayer2006, Kim2010a, Janssen2015a} The hope, as in previous decompositions, is that the correction terms are small or can be obtained with fewer computational resources than would otherwise be required to directly calculate the original multitime correlation function.

Applying the same method to the $N$-time propagator,
\begin{equation}\label{eq:ntimeprop}
\begin{split}
    \boldsymbol{\mc{C}}^{B_{N-1},...,B_1}&(t_N,...,t_2,t_1) = \\
    &(\mathbf{A}|\e{\mc{L}t_N}B_{N-1}...\e{\mc{L}t_2} B_1 \e{\mc{L}t_1}|\mathbf{A}),
\end{split}
\end{equation}
gives rise to higher-order memory kernel terms,\cite{Ivanov2015a}
\begin{equation}\label{eq:mem_general}
\begin{split}
    &\bm{\mc{K}}^{B_{N-1},...,B_1}(t_N,...,t_2,t_1) = \\
    &\qquad (\mathbf{A}|\mc{L} \e{\mc{QL}t_N}  \mc{Q} B_{N-1} ...\e{\mc{L}t_2}\mc{Q}B_1 \e{\mc{LQ}t_1} \mc{QL} |\mathbf{A}).
\end{split}
\end{equation}
For the 3-time propagator required for 4-wave mixing spectroscopies, the expression is
\begin{widetext}
\begin{equation}\label{eq:3timeprop}
\begin{split}
    \bm{\mc{C}}_{(t_3, t_2, t_1)}^{B_2,B_1} &= \bm{\mc{C}}(t_3) \bm{\mc{B}}_2 \bm{\mc{C}}(t_2) \bm{\mc{B}}_1 \bm{\mc{C}}(t_1) 
                                    +\bm{\mc{C}}(t_3) \bm{\mc{B}}_2 \bm{\mc{C}}(t_2-\tau_2) \bm{\mc{K}}^{B_1}_{\tau_2, \tau_1} \bm{\mc{C}}(t_1-\tau_1) 
                                     + \bm{\mc{C}}(t_3-\tau_3) \bm{\mc{K}}^{B_2}_{\tau_3, \tau_2} \bm{\mc{C}}(t_2-\tau_2) \bm{\mc{B}}_1 \bm{\mc{C}}(t_1) \\
                                    &~~~ + \bm{\mc{C}}(t_3-\tau_3) \bm{\mc{K}}^{B_2}_{\tau_3, \tau_2} \bm{\mc{C}}(t_2-\tau_2-\tau_2') \bm{\mc{K}}^{B_1}_{\tau_2', \tau_1} \bm{\mc{C}}(t_1-\tau_1) 
                                     + \bm{\mc{C}}(t_3-\tau_3) \bm{\mc{K}}_{\tau_3, t_2, \tau_1}^{B_2, B_1} \bm{\mc{C}}(t_1-\tau_1),
\end{split}
\end{equation}
\end{widetext}
where the identities of $B_1, B_2$ depend on the type of spectroscopy. For example, in 2D electronic spectroscopy $B_2=\mu_\mathrm{R}^\times$, the commutator with the raising transition dipole moment operator, and $B_1 = \mu_\mathrm{L}^\times$ for the non-rephasing contribution, while $B_1=B_2$ in the rephasing case.\cite{Mukamel1999a} As in the 2-time case, the expression contains terms with differing amounts of temporal correlation: the final term describes 3-time correlations, which manifest as the 3-time kernel $\bm{\mc{K}}_{t_3, t_2, t_1}^{B_2, B_1}$; the middle three terms describe the 2-time correlations described in Eqs.~\ref{eq:2timeprop_decomp}~and~\ref{eq:2timemem}; and the fully 1-time reducible, homogeneous evolution is given by the leading $\Pi_{i=1}^{n-1}\left(\mc{P}B_i\mc{P}\right)$ term in the decomposition from substituting Eq.~\ref{eq:system_split} into Eq.~\ref{eq:ntimeprop} at $3^\rmm{rd}$ order. For practical purposes, Eq.~\ref{eq:2timeprop_decomp} can be used to condense the first four terms into just two convolutions (see Eq.~\ref{eq:U3_condensed}), which reduces computer memory requirements when employing these decompositions.

The fact that the 3-time contribution does not contain a convolution integral over the $t_2$ axis is qualitatively different from the 1- and 2-time expressions in Eqs.~\ref{eq:NZ}~and~\ref{eq:2timeprop_decomp}. For example, in the MNZ equation (Eq.~\ref{eq:NZ}), $\bm{\mc{K}}(t>\tau_\mc{K}) = \bm{0}$ implies that one has entered into a regime where the evolution of the low dimensional set of observables does not require knowledge of its history beyond a certain memory time. Procedurally, the need for an integration remains, but the window of the convolution is now of fixed length: The term containing $\bm{\mc{K}}(t)$ is still required to evaluate Eq.~\ref{eq:NZ} at $t>\tau_\mc{K}$. In contrast, for the 3-time term $\bm{\mc{K}}^{B_1, B_2}(\tau_3, t_2 > t_{2, \rmm{max}}, \tau_1) = \bm{0}$ means that the \textit{entire contribution} goes to zero: The term can be discarded. This observation has a simple physical interpretation: when $t_2$ is sufficiently large, the $t_1$ and $t_3$ regions become fully decorrelated from one another. At this point, calculation of \textit{any} 3-time correlation function has, at most, the cost of a 2-time correlation function. 

The clear and compact disentanglement of multitime correlations into terms of increasing correlation (and therefore computational expense to obtain) offers distinct advantages. In viewing the problem as corrections to the homogeneous evolution that ignores multitime correlations, statements like that in the previous paragraph (concerning the lifetime of each contribution with respect to the $\tau_\mc{K}$ appropriate for the 1-time MNZ kernel) can be made and exploited to yield both physical insight and algorithmic efficiency. Indeed, anticipating our results, in some regimes, higher-order correlations become merely small perturbations, while in others they are the source of the correct long-time physics. Unfortunately, direct application of this method as is can still be intensely computationally demanding. A 2D spectrum, for example, is a three-index array that can be seen as a list of images (the spectrum at time $t_2$). The use of the Dyson identity to derive the memory kernels demands numerical derivatives and therefore a small underlying time step along each time axis to avoid discretization errors. Then, the Fourier transform requires extending the arrays (along $t_1$ and $t_3$) until the response functions have decayed. In combination, this leads to very large objects that demand high costs in both time and space, especially when employing numerically exact methods to generate the reference dynamics. It is these costs that motivate us to search for a more parsimonious implementation.

\vspace{-4pt}
\subsection{Discrete time transfer-tensor construction}\label{sec:TTM_theory}
\vspace{-6pt}

The transfer tensor method decomposes a non-Markovian evolution using layered dynamical maps. Specifically, one systematically unravels the dynamics by isolating the deviation from Markovianity at each time slice as a sum of 1-, 2-, ..., $N$-time entanglements with the previous states of the system.\cite{Cerrillo2014a} Formally, the transfer-tensor is the object $\bm{\mc{T}}_n$ that satisfies the discrete convolution 
\begin{equation}\label{eq:discreteConv}
    \bm{\mc{C}}(N\delta t) = \sum_{j=0}^{N-1} \bm{\mc{T}}_{N-j} \bm{\mc{C}}(j\delta t),
\end{equation}
where the underlying dynamics from which $\bm{\mc{T}}_n$ is constructed usually have a quasi-continuous, higher resolution than the $\delta t$ of the transfer tensor. Information at time steps between these longer time `slices' is effectively \textit{discarded and cannot be recovered later}. This means that the temporal resolution of dynamics predicted using the transfer tensor is limited by the size of the time steps used to construct the tensor. The fact that we have written $\bm{\mc{T}}_n$ with only a single index implies time-translational invariance and thus the lack of an inhomogeneous term in the MNZ equation, Eq.~\ref{eq:NZ}. Like $\bm{\mc{K}}(t)$, $\bm{\mc{T}}_n$ goes to zero in finite time in dissipative systems, and so can be obtained in full from sufficiently long reference dynamics as
\begin{equation}\label{eq:TTM_def}
    \bm{\mc{T}}_N = \left[ \bm{\mc{C}}(N\delta t) - \sum_{j=1}^{N-1} \bm{\mc{T}}_{N-j} \bm{\mc{C}}(j\delta t) \right]\bm{\mc{C}}^{-1}(0) ,
\end{equation}
where the dependence on $\bm{\mc{C}}^{-1}(0)=\mathbb{1}$ can be dropped. One can construct the tensor iteratively, with increasing $N=1, 2, ...$, until $\bm{\mc{T}}_{n}=\bm{0}~\forall~n\geq n_\rmm{max}$. 

An important feature of the TTM is that it circumvents time discretization errors by not requiring time derivatives for its construction. This property can be further exploited, as it gives us a route to massively increase the sparseness of the data required to perform projected equation of motion methods. This is advantageous in light of the steep computational cost of the Ivanov-Breuer decomposition described in Sec.~\ref{sec:multi_theory}. However, direct application of the TTM in the presence of an inhomogeneous term, which invariably arises in multitime cases, causes the description to become complicated.\cite{Jørgensen2020a, Gherardini2022} In existing TTM-based approaches, the desirable separability of the MNZ into two terms is not easily reproduced, as the initial condition dependence is subsumed into the overall convolutional structure. Direct TTM routes to solving the multitime problem therefore do not offer the same separability and consequently sacrifice physical transparency.

\subsection{Discrete-time formulation of multitime memory kernel theory}

Although the TTM can be motivated in terms of layered dynamical maps, it is also the discrete form of the integrated MNZ equation. The relationship with $\bm{\mc{K}}(t)$ can be seen by taking a finite difference of Eq.~\ref{eq:discreteConv},\cite{Pollock2018b}
\begin{align}\label{eq:TcomparedtoK}
    \frac{\bm{\mc{C}}(N\delta t) - \bm{\mc{C}}\big((N-1) \delta t\big)}{\delta t} &= \frac{\bm{\mc{T}}_1 - \mathbb{1}}{\delta t}\bm{\mc{C}}\big((N-1)\delta t\big) \nonumber\\
    &~~~~~+\sum_{j=0}^{N-1} \delta t \frac{\bm{\mc{T}}_{N-j}}{\delta t^2}\bm{\mc{C}}(j\delta t)
\end{align}
and noting that since $\bm{\mc{T}}_1 = \bm{\mc{C}}(\delta t)$, in the limit of $\delta t \rightarrow 0$ this Eq.~\ref{eq:TcomparedtoK} becomes the inhomogeneity-free Eq.~\ref{eq:NZ} when $\bm{\mc{C}}(0)=\mathbb{1}$. Therefore one can identify
\begin{equation}\label{eq:deltat_facs}
    \bm{\mc{K}}(N \delta t) = \lim_{\delta t\rightarrow 0}{\bm{\mc{T}}_{N}/\delta t^2}.
\end{equation} 
This defines $\tau_\mc{K} = n_\rmm{max}\delta t$. While $\bm{\mc{T}}_n$ increasingly deviates from the memory kernel $\bm{\mc{K}}(t)$ as the time slice becomes coarser (see Eq.~\ref{eq:deltat_facs} and accompanying discussion, below), the equations they satisfy result in equivalent dynamics. As such, a TTM-like approach yields the required memory kernels and converts them to $\bm{\mc{T}}$-like objects that exist in discrete time. What is more, by not requiring the evaluation of auxiliary kernels (see Appendix~\ref{app:auxiliary_forms}), this approach eliminates a number of computational steps. Of course, similar iterative solutions have been used in the context of 1-time correlation functions without reference to the quantum dynamics problem.\cite{Berkowitz1981, Kowalik2019}

The key insight we propose to exploit is that, from a purely numerical standpoint, \textit{this routine allows one to decompose any time series as a discrete convolution against itself}. Moreover, as a numerical method, one can go further than a self-convolution, as there is no practical reason that the time series on the right hand side of Eq.~\ref{eq:discreteConv} need be the same as that on the left hand side for an expression like Eq.~\ref{eq:TTM_def} to follow.\cite{Berkowitz1981} The requirement is simply that this object is already known. As an equation,
\begin{equation}\label{eq:discreteConvAB}
    \bm{\mc{E}}_N = \sum_{j=0}^{N-1} \bm{\mc{F}}_{N-j} \bm{\mc{G}}_j,
\end{equation}
can be iteratively solved for 
\begin{equation}\label{eq:TTM_defAB}
    \bm{\mc{F}}_N = \left[ \bm{\mc{E}}_N - \sum_{j=1}^{N-1} \bm{\mc{F}}_{N-j} \bm{\mc{G}}_j \right]\bm{\mc{G}}^{-1}_0 ,
\end{equation}
provided the functions $\bm{\mc{E}}_j$ and $\bm{\mc{G}}_j$ are known. By careful, repeated application, this simple change is all that is required to bring Eq.~\ref{eq:3timeprop} to discrete time, which maintains the clear, physically transparent interpretation of the memory terms. We demonstrate in detail how we achieve this in the next section.

\vspace{-4pt}
\section{Discrete Multitime Propagation}
\vspace{-6pt}

In what follows we consider two models. We employ the numerically exact Hierarchical Equations of Motion (HEOM) method\cite{Tanimura1989, Ishizaki2005, Xu2007} to generate all reference dynamics (see Appendix~\ref{app:comp}). In Sec.~\ref{sec:spin-boson_results}, we employ the spin-boson model to demonstrate that our proposed mathematical framework functions as we have outlined, starting with the 2-time decomposition of Eq.~\ref{eq:2timeprop_decomp}, shown in Fig.~\ref{fig:2timeprop}, and then in the 3-time decomposition of Eq.~\ref{eq:3timeprop}, shown in Fig.~\ref{fig:3timeprop}. We choose this minimal model for reasons of efficiency and ease of visualization. In Sec.~\ref{sec:eet_results}, we move to the EET dimer model, which offers the smallest relevant model in electronic 2D spectroscopy. We show two different parameter regimes: one in the limit of weak exciton-phonon coupling ($\lambda \ll \Delta$) and another with intermediary coupling ($\lambda = \Delta / 2$) strength, shown in  Figs.~\ref{fig:underdamped}~and~\ref{fig:damped}, respectively. In the weak coupling case, we showcase the ability of our method to extend the response functions along their ($t_1$ and $t_3$) axes to increase spectral resolution in long-lived, underdamped cases. We also show how the 2-time and 3-time contributions to the full spectrum are both small compared to the homogeneous, 1-time term. In the intermediate coupling case, we encounter a system where the 2-time and 3-time contributions are critically important to the form of the spectrum. We also extend the spectra along the waiting time ($t_2$) axis, and confirm that the 3-time contribution is negligible at long $t_2$ times in Fig.~\ref{fig:extend_t2}.

\subsection{Spin-boson validation}\label{sec:spin-boson_results}
\vspace{-6pt}

We choose $B_1 = \sigma_x$ and $B_2 = \sigma_y$. Anticipating the need to store up to 5-index arrays, we choose a parameter regime known to have a fast-decaying memory kernel such that, when combined with the minimal Hilbert space of the 2-state system, it leads to a data set with only moderate computational cost. This will allow us to perform the original unraveling in quasi-continuous time and compare it to the output of our new approach.

\subsubsection{2-time propagator}\label{ssec:spin-boson_results_2time}
\vspace{-6pt}

We first obtain $\bm{\mc{T}}^B_{t_2, N\delta t_1}$, the discrete-time version of $\bm{\mc{K}}^B(t_2, t_1)$, from a TTM-like decomposition of Eq.~\ref{eq:2timeprop_decomp}. We then check that this 1-measurement memory object returns the 2-time propagator, as it must by construction, and compare its structure with the result obtained from the direct unraveling of the continuous-time definition (Eq.~\ref{eq:2timemem}, see Appendix~\ref{app:auxiliary_forms} for details). The requisite 1- and 2-time propagators are the direct output of the HEOM simulation.

For this demonstration, we discretize only the second time index of Eq.~\ref{eq:2timeprop_decomp} such that $\delta t_2 = 10 \delta t_1$, with $t_1$ remaining quasi-continuous. When quantities are available only at these discrete time slices we write the relevant indices as subscripts, for clarity. We can access the 2-time discrete memory kernel in Eq.~\ref{eq:2timeprop_decomp} by rearranging it as
\begin{align}
    \bm{\mathfrak{L}}^B_{t_2}(t_1) &\equiv \bm{\mathcal{C}}^B_{t_2}(t_1) - \bm{\mc{C}}_{t_2}\bm{\mc{BU}}(t_1)\nonumber\\
    &= \bm{\mc{C}}_{t_2-\tau_2}\bm{\mc{K}}^B_{t_2-\tau_2}(t_1-\tau_1)\bm{\mc{C}}(t_1-\tau_1) \label{eq:2timeLHS_def}\\
    &= \sum_{\tau_2}\bm{\mc{C}}_{t_2-\tau_2}\int_0^{t_1}\dd{\tau_1\,}\bm{\mc{K}}^B_{\tau_2}(\tau_1)\bm{\mc{C}}(t_1-\tau_1),\label{eq:2time_TTM_1}
\end{align}
where the third line explicitly shows the sums and integrals. The key point is that $\bm{\mathfrak{L}}^B_{t_2}(t_1)$ can be constructed from the exact simulation: $\bm{\mc{C}}^B_{t_2}(t_1)$ and $\bm{\mc{C}}(t_1)$ are the direct output of the simulation and we obtain $\bm{\mc{C}}_{t_2-\tau_2}$ by sub-sampling $\bm{\mc{C}}(t_1)$ every $10$ points. We then rewrite Eq.~\ref{eq:2timeLHS_def} as 
\begin{equation}
    \bm{\mathfrak{L}}^B_{t_2}(t_1) = \sum_{\tau_2}\bm{\mc{C}}_{t_2-\tau_2}\bm{\mathfrak{T}}^B_{\tau_2}(t_1)
\end{equation}
where the convolution of the 2-time memory kernel and the 1-time propagator yield the unknown,
\begin{equation}\label{eq:mathfrak_T_2time}
    \bm{\mathfrak{T}}^B_{\tau_2}(t_1) \equiv \int_0^{t_1}\dd{\tau_1\,}\bm{\mc{K}}^B_{\tau_2}(\tau_1)\bm{\mc{C}}(t_1-\tau_1).
\end{equation}
In analogy to extracting $\bm{\mc{T}}_j$, but for the mixed version with the form of Eq.~\ref{eq:TTM_defAB}, we can write the term containing the memory kernel as an iteratively evaluated expression
\begin{equation}\label{eq:2time_mathfrak_ttm}
    \bm{\mathfrak{T}}^B_{t_2}(t_1) = \bm{\mathfrak{L}}^B_{t_2}(t_1) - \sum_{j=1}^{N-1} \bm{\mathfrak{T}}^B_{j\delta t_2}(t_1) \bm{\mc{C}}_{t_2-j\delta t_2}.
\end{equation}

We can then extract the 2-time memory kernel, $\bm{\mc{K}}^B_{\tau_2}(\tau_1)$, from Eq.~\ref{eq:mathfrak_T_2time}. Although we have not coarsened the $t_1$ axis, we can apply a second discretization to deconvolve the integral in Eq.~\ref{eq:mathfrak_T_2time}, which is equivalent to using the trapezium rule to obtain $\bm{\mc{K}}(t)$ from auxiliary kernels.\cite{Pfalzgraff2019a} Converting to a discrete sum defines the doubly-discrete analogue to $\bm{\mc{K}}^{B}(t_2, t_1)$ as $\bm{\mc{T}}^B_{t_2,j\delta t_1}$, which satisfies
\begin{equation}
    \bm{\mathfrak{T}}^B_{\tau_2}(t_1) = \sum_{j=0}^{N-1} \bm{\mathcal{T}}^B_{t_2, j\delta t_1} \bm{\mc{C}}(t_1 - j\delta t_1).
\end{equation}
As in Eq.~\ref{eq:2time_mathfrak_ttm}, we can rearrange this expression (which again has the form of Eq.~\ref{eq:TTM_defAB}) to extract the discrete, 2-time transfer-matrix,
\begin{equation}\label{eq:2time_TTM_2}
    \bm{\mc{T}}^B_{t_2, N\delta t_1} = \bm{\mathfrak{T}}^B_{t_2}(N\delta t_1) - \sum_{j=1}^{N-1} \bm{\mathcal{T}}^B_{t_2, j\delta t_1} \bm{\mc{C}}(t_1-j\delta t_1).
\end{equation}

\begin{figure}[!t]
\begin{center} 
    \vspace{-10pt}
    \resizebox{.45\textwidth}{!}{\includegraphics{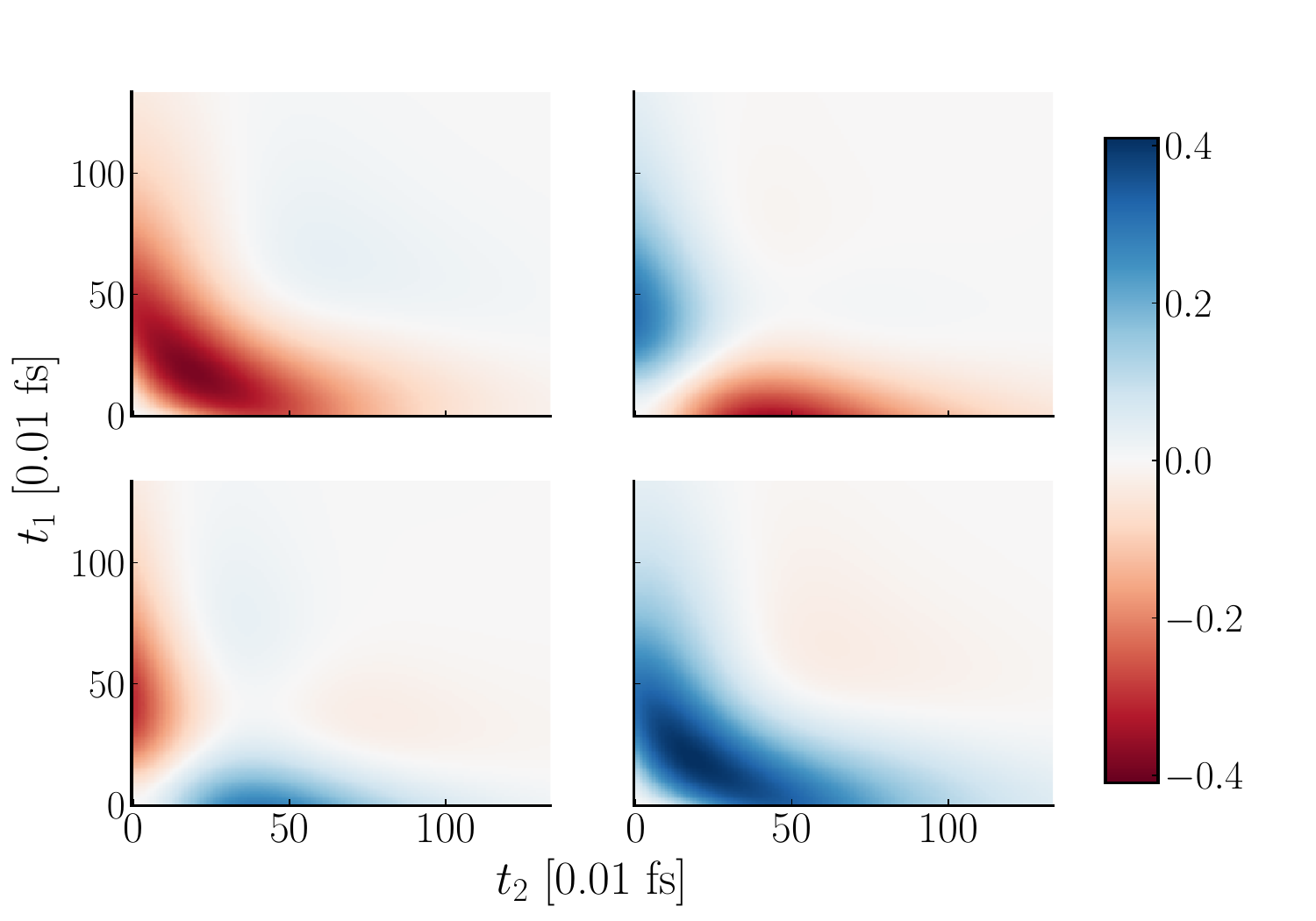}}
    \\
    \resizebox{.45\textwidth}{!}{\includegraphics[trim={0 0 0 0.5cm},clip]{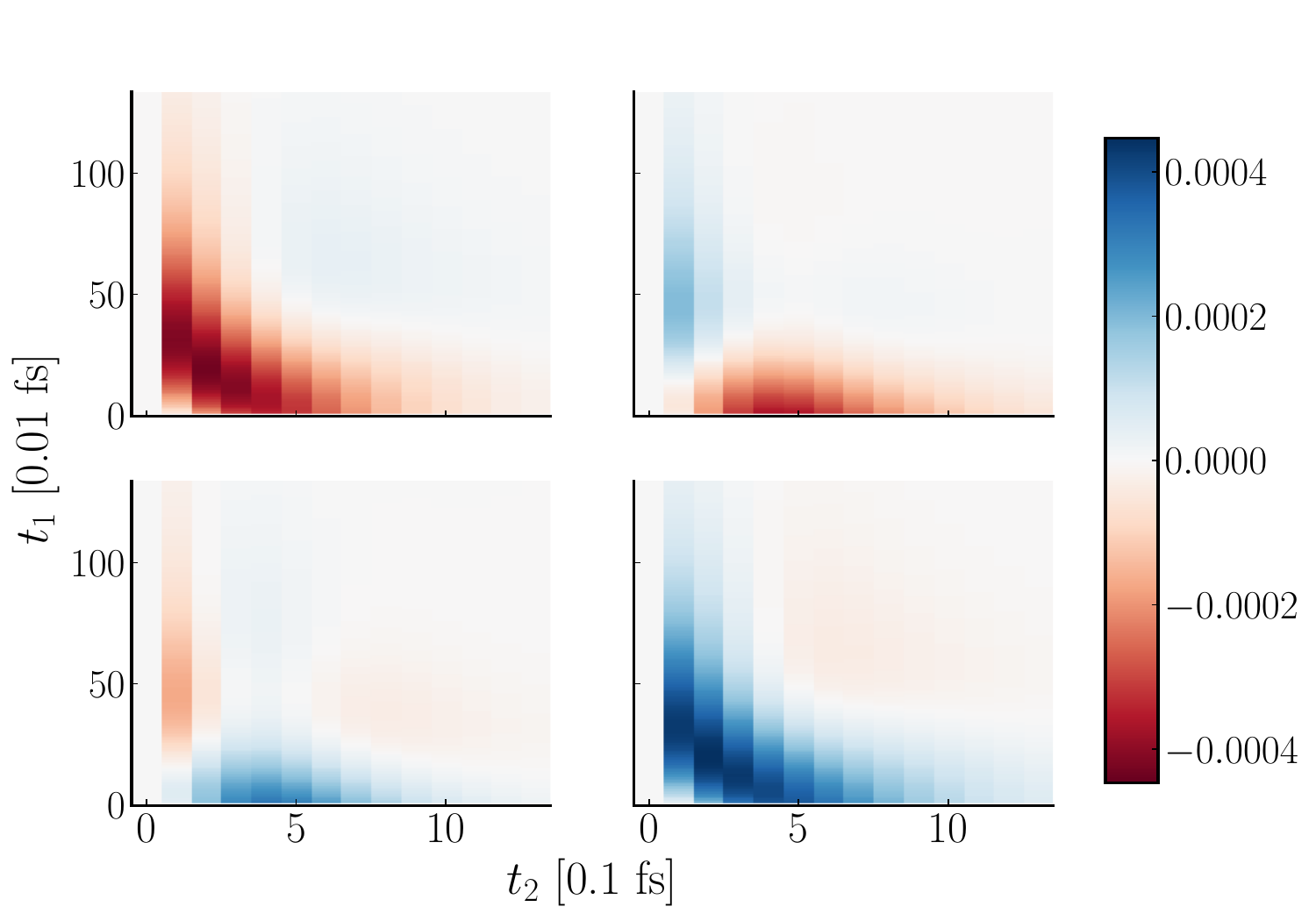}}
    \vspace{-5pt}
\caption{\label{fig:2timeprop} Center 2x2 elements of the larger 4x4 2-time kernels for the spin-boson model. They have a banded structure where only the 2$^\rmm{nd}$ and 3$^\rmm{rd}$ rows are non-zero, and the 1$^\rmm{st}$ and 4$^\rmm{th}$ columns of those rows are negligible in size. \textbf{Top}: $\bm{\mc{K}}^B$ of Eq.~\ref{eq:2timemem}. \textbf{Bottom}: $\bm{\mc{T}}^B$ of Eq.~\ref{eq:2time_TTM_2}. The first discrete step in $t_2$ is null because the 2-time propagator is completely described by the homogeneous term up to that point, meaning $\bm{\mathfrak{L}}^B_0(t_1) = \bm{0}$. The difference in magnitude between the two objects is due to the two TTM steps: the first scales by a factor of $\delta t_2$, and the second by a factor of $\delta t_1$, leading to an overall factor $1000$ reduction in magnitude.}
\end{center}
\vspace{-20pt}
\end{figure}

Figure~\ref{fig:2timeprop} compares $\bm{\mathcal{T}}^B_{t_2, N\delta t_1}$ with $\bm{\mc{K}}^B(t_2, t_1)$ obtained by the previous, continuous time method (see Appendix~\ref{app:auxiliary_forms}). Our discretization method performs perfectly, as both objects offer an exact rewriting of the exact 2-time propagator, albeit in discrete and continuous time (Eqs.~\ref{eq:2timeprop_decomp}~and~\ref{eq:2timeLHS_def}) respectively.

\vspace{-4pt}
\subsubsection{3-time propagator}\label{ssec:spin-boson_results_3time}
\vspace{-6pt}
Now we extend the approach to obtain the 3-time propagator. The basic idea is the same as before. We isolate the unknown $\mc{T}_{t_3,t_2,t_1}^{B_1, B_2}$, the discrete-time analogue to $\mc{K}^{B_2, B_1}(t_3, t_2, t_1)$, on the right hand side of a rearranged and discretized Eq.~\ref{eq:3timeprop}. The quantity constructed from the reference dynamics is
\begin{align}\label{eq:rearrange_3timeprop}
    \bm{\mathfrak{L}}^{B_2, B_1}_{t_3, t_2, t_1} &\equiv \bm{\mc{C}}_{t_3, t_2, t_1}^{B_2,B_1} - \bm{\mc{C}}_{t_3} \bm{\mc{B}}_2 \bm{\mc{C}}^{B_1}_{t_2, t_1} - \bm{\mc{C}}_{t_3-\tau_3}\bm{\mc{T}}^{B_2}_{\tau_3, \tau_2}\bm{\mc{C}}^{B_1}_{t_2-\tau_2, t_1} \nonumber\\
                                    &= \bm{\mc{C}}_{t_3-\tau_3} \bm{\mc{T}}_{\tau_3, t_2, \tau_1}^{B_2, B_1} \bm{\mc{C}}_{t_1-\tau_1},
\end{align}
where the subscript notation signals that we intend to work with discrete data on all indices from now on. We omit the sums over repeated indices for clarity. The 2-time propagator, $\bm{\mc{C}}^{B_1}_{t_2,t_1}$ is obtained from the same simulation that generates $\bm{\mc{C}}_{t_3, t_2, t_1}^{B_2,B_1}$, along with $\bm{\mc{C}}_{t}$. Note however that the resolution along the $t_1$ and $t_3$ axes will be finer than that along $t_2$ since we intend to perform Fourier transforms over those axes to obtain 2D spectra. As such, computation of $\bm{\mc{T}}^{B_2}_{t_3, t_2}$ is slightly complicated as it assumes the longer time-slice on the first index instead of the second. This means we have to perform a separate HEOM simulation to acquire this 2-time kernel, as it is not contained in the data used to construct the 3-time propagator. The reason for this is that there is no general way to remove the effect of the $B_1$ measurement, even when $t_1=0$. Fortunately, since a 2-time calculation is orders of magnitude cheaper than the 3-time calculation itself, this is not a significant cost. Care must be taken to keep track of factors of $\delta t$ when combining terms with different numbers of TTM steps, as each conversion to a discrete sum is a unit conversion accompanied by a relation like Eq.~\ref{eq:deltat_facs}. Once the additional data has been generated, $\bm{\mc{T}}^{B_2}_{t_3, t_2}$ is obtained exactly as described in the previous section.

\begin{figure}[!t]
\vspace{-12pt}
\begin{center} 
    \hspace{-12pt}
    \resizebox{.25\textwidth}{!}{\includegraphics{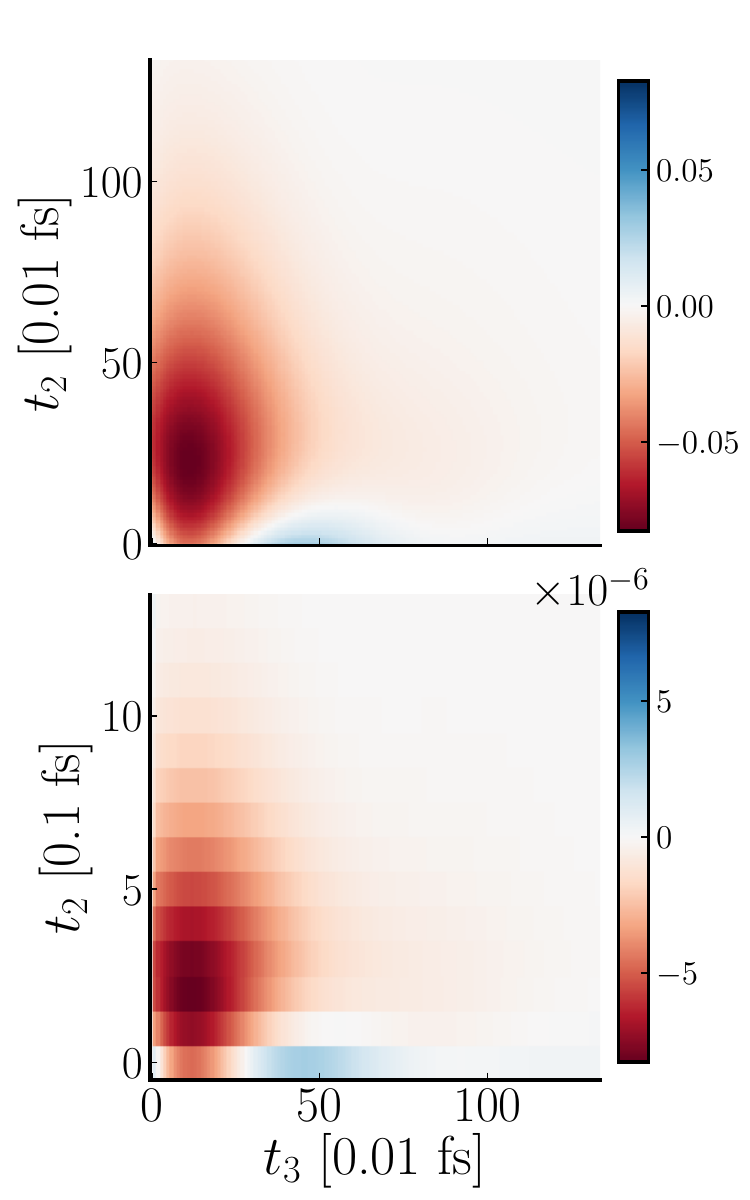}}
    \hspace{5pt}
    \resizebox{.15\textwidth}{!}{\includegraphics{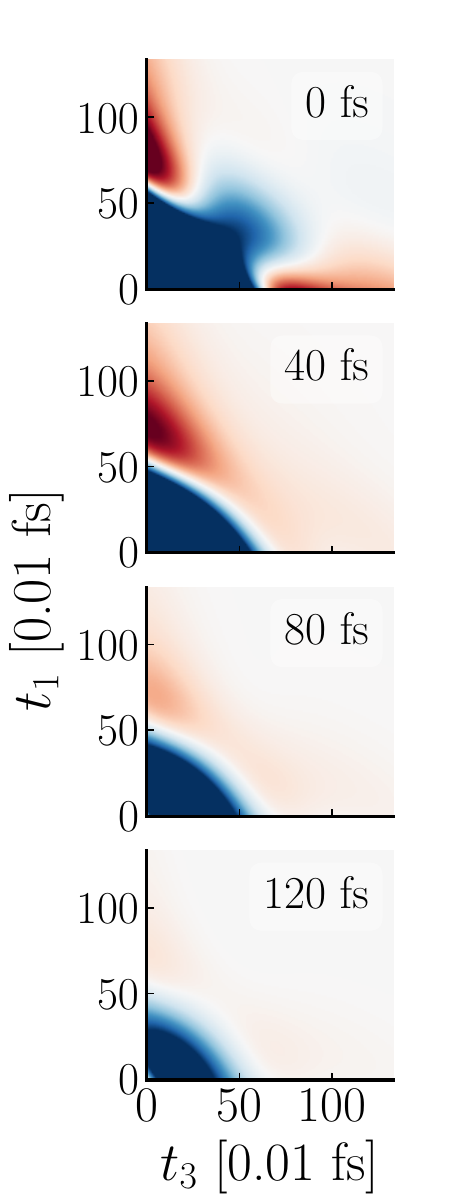}}
\caption{\label{fig:3timeprop} 3-time kernel elements $\mc{K}_{1,1}^{B_1,B_2}$ and $\mc{T}_{1,1}^{B_1,B_2}$ for the spin-boson model. \textbf{Top Left}: $\mc{K}^{B_2,B_1}$ of Eq.~\ref{eq:3timemem} $t_1 = 80 * \delta t_1 = 0.8$~fs, which is the time the $B_1$ measurement is made. \textbf{Bottom Left}: $\mc{T}^{B_2,B_1}$ of Eq.~\ref{eq:rearrange_3timeprop} at the same $t_1$ time. The TTM steps are both $\delta t_1$ in size, totaling a factor $10000$ reduction in magnitude compared to the continuous case. \textbf{Right}: The three time kernel in the $t_1, t_3$ plane at increasing $t_2$ values; color-scale the same as on the left.}
\end{center}
\vspace{-10pt}
\end{figure}

As in Eqs.~\ref{eq:2time_TTM_1}--\ref{eq:2time_TTM_2}, we form two consecutive TTM-like equations, each with $\bm{\mc{C}}_t$ under the convolution, once on the left and once on the right. Both are sums over the shorter time slices, and we loop over the longer $t_2$ index. While this might make this step relatively expensive, as we noted at the end of Sec.~\ref{sec:multi_theory}, the 3-time contribution is only required for the values of $t_2 \leq t_{2,\rmm{max}}$ one intends to analyze. Again, the system is small enough that the fully continuous $\bm{\mc{K}}^{B_2, B_1}(t_3, t_2, t_1)$ can be unraveled and compared to $\bm{\mc{T}}_{t_3,t_2,t_1}^{B_1, B_2}$ with one of the time-indices fixed to allow for a 2D visualization. Figure~\ref{fig:3timeprop} compares the continuous- and discrete-time 3-time kernels for a given $t_1$ value, showing clear agreement.

As an additional check on the method, one can consider the limit of no waiting time between measurements when $t_2 = 0$. In such cases, the 3-time correlation function reduces to a 2-time problem. The structure of the 3-time kernel from Eq.~\ref{eq:mem_general} shows that
\begin{equation}\label{eq:3timemem}
    \bm{\mc{K}}_{t_3, t_2, t_1}^{B_2, B_1} = (\mathbf{A}| \mc{L}\e{\mc{QL}t_3} \mc{Q} B_2 \e{\mc{QL}t_2} \mc{Q} B_1 \e{\mc{QL}t_1} \mc{Q} \mc{L} |\mathbf{A}), 
\end{equation}
and therefore,
\begin{align}
    \bm{\mc{K}}_{t_3, 0, t_1}^{B_2, B_1} &= (\mathbf{A}| \mc{L}\e{\mc{QL}t_3} \mc{Q} \left(B_2 \mc{Q} B_1\right) \e{\mc{QL}t_1} \mc{Q} \mc{L} |\mathbf{A}) \nonumber\\
        &= (\mathbf{A}| \mc{L}\e{\mc{QL}t_3} \mc{Q} \left(B_2 B_1\right) \e{\mc{QL}t_1} \mc{Q} \mc{L} |\mathbf{A})   \nonumber\\
           &~~ - (\mathbf{A}| \mc{L}\e{\mc{QL}t_3} \mc{Q} B_2 |\mathbf{A}) (\mathbf{A}| B_1 \e{\mc{QL}t_1} \mc{Q} \mc{L} |\mathbf{A}). \label{eq:BQB_terms}
\end{align}
When $B_2$ belongs to the projector, $|\mathbf{A})$, the second term in Eq.~\ref{eq:BQB_terms} that contains $\mc{Q}B_2|\mathbf{A})$ vanishes: $B_2|\mathbf{A})$ is still in $\mc{P}$, so it returns zero when $\mc{Q}$ acts on it. The same is true of the $(\mathbf{A}|B_1\e{\mc{QL}t_1}\mc{Q}$ part if $B_1$ belongs to $(\mathbf{A}|$, since we can commute the $\mc{Q}$ through the projected propagator. As a result, the 3-time kernel with $t_2=0$ is equal to a 2-time kernel for a measurement of $B_2B_1$. That is,
\begin{equation}
        \bm{\mc{K}}_{t_3, 0, t_1}^{B_2, B_1} = (\mathbf{A}| \mc{L}\e{\mc{QL}t_3} \mc{Q} \left(B_2 B_1\right) \e{\mc{QL}t_1} \mc{Q} \mc{L} |\mathbf{A}) \equiv \bm{\mc{K}}_{t_3, t_1}^{(B_2 B_1)}.
\end{equation}
In our case $B_2B_1 = -i\sigma_z$. Running the corresponding 2-time simulation numerically confirms the equality with the 3-time kernel at $t_2=0$.

\begin{figure*}[!t]
\begin{center} 
    \hspace{-15pt}
    \resizebox{.3\textwidth}{!}{\includegraphics{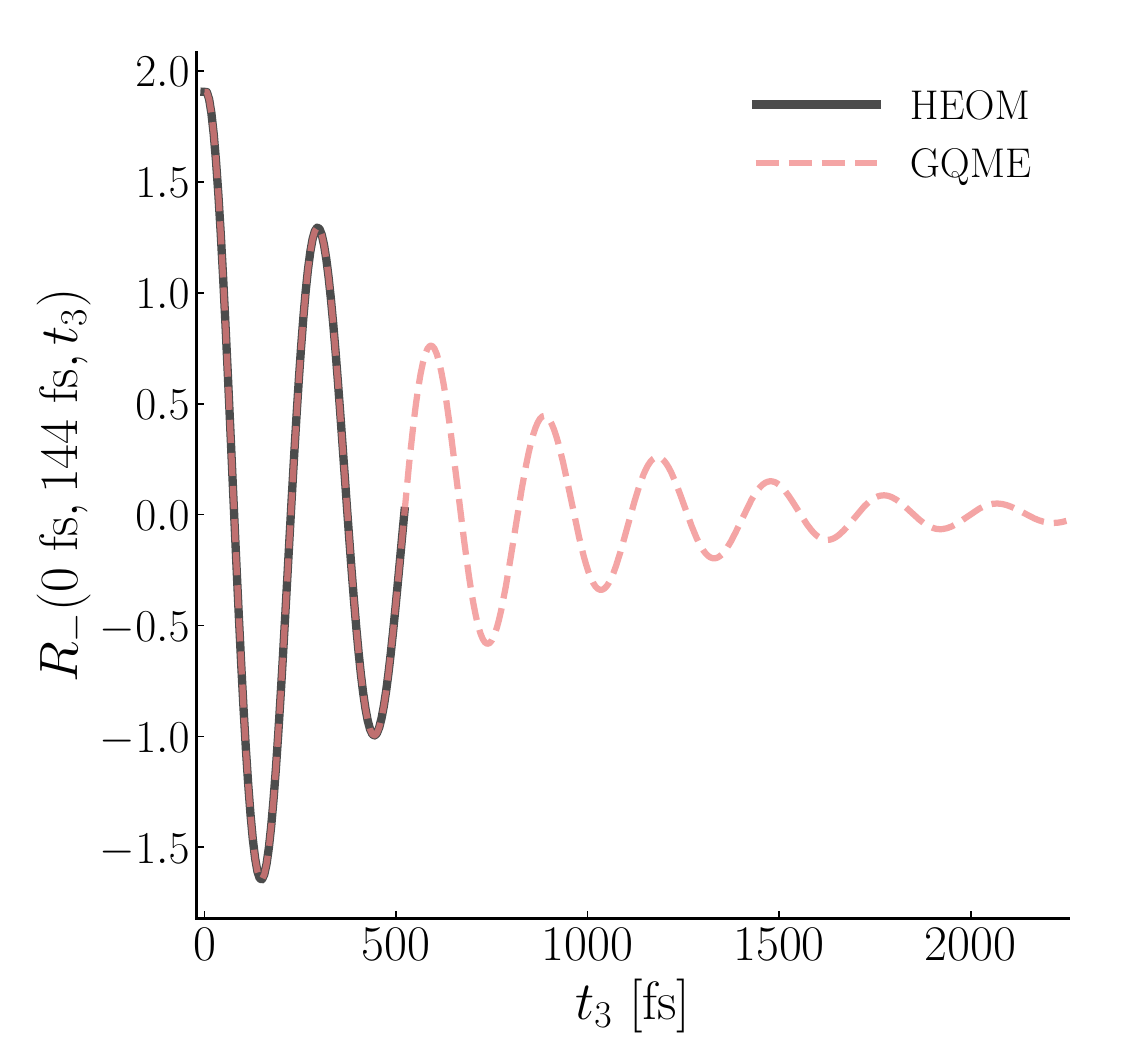}}\hspace{-1pt}
    \resizebox{.4\textwidth}{!}{\includegraphics{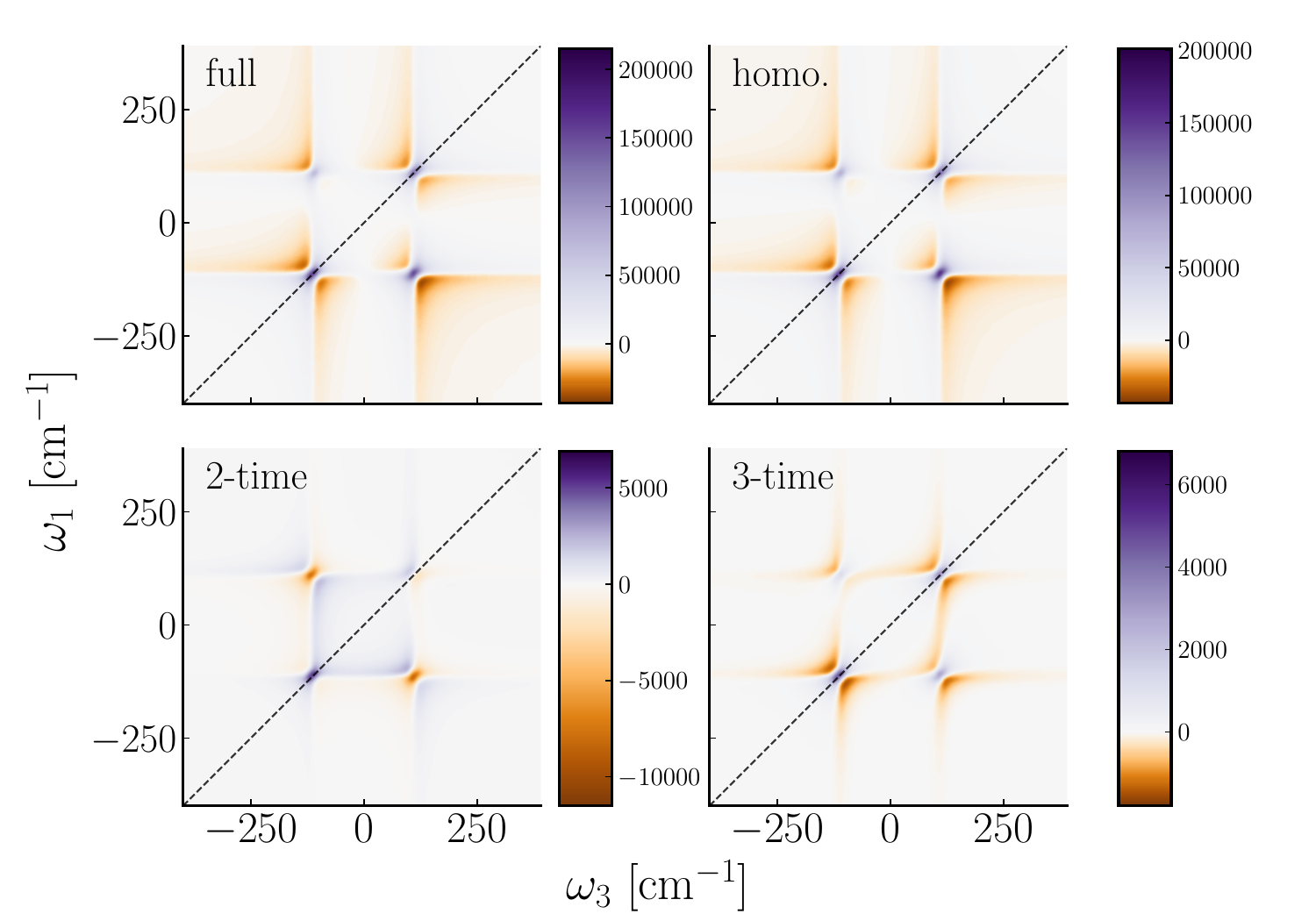}}\hspace{-1pt}
    \resizebox{.2\textwidth}{!}{\includegraphics{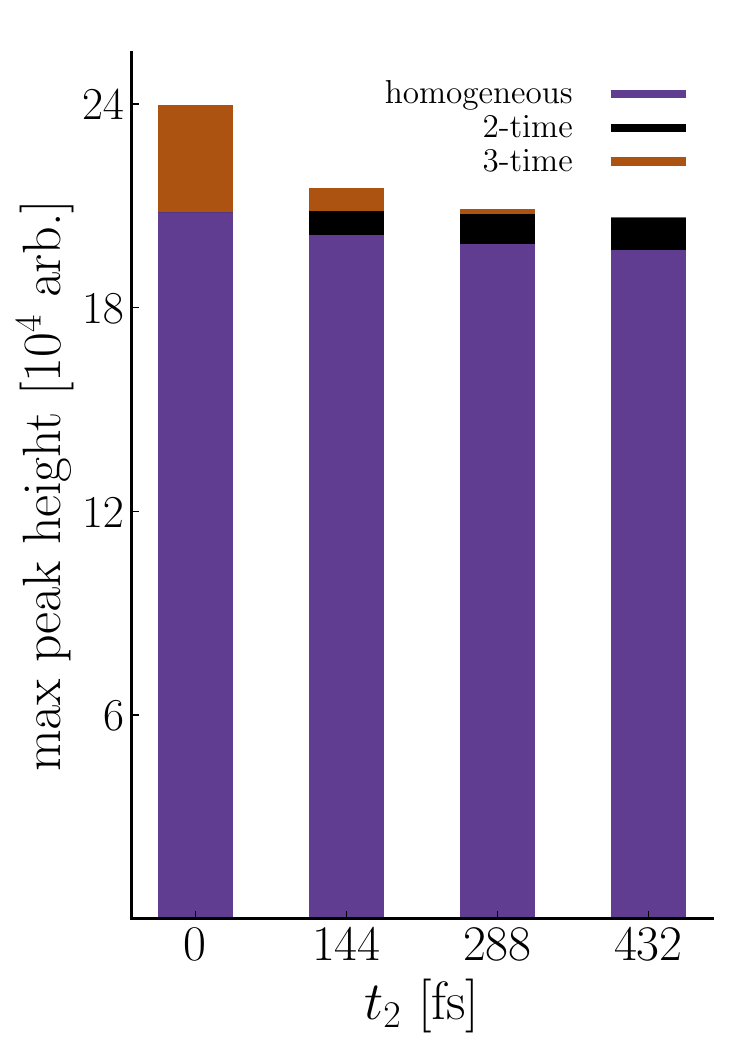}}
\end{center}
\vspace{-16pt}
\caption{\label{fig:underdamped} EET dimer with $\pm\epsilon = \pm50~\rmm{cm}^{-1}$ and $\Delta = 100~\rmm{cm}^{-1}$, while $\lambda = 2~\rmm{cm}^{-1}$; $\tau = 100~\mathrm{fs}$ and $T=298$~K, $K=0$ and $L=5$. \textbf{Left}: $R_-$ at a representative $t_1$ and $t_2$ time. The HEOM (solid black) results are not long enough to converge the response function for use in Eq.~\ref{eq:Iphasing}, but the GQME method described in the text allows it to be both reconstructed and extended (dashed red) so that resolved spectra may be obtained. \textbf{Middle}: Decomposition of the 2D spectrum at $t_2=144$~fs. Most of the spectrum is captured by the homogeneous term. The 2-time contributions have a different structure to the 1- and 3-time counterparts. \textbf{Right}: Decomposed contributions' maxima for each of the $t_2$ times. The homogeneous term is relatively insensitive to $t_2$. As expected, 3-time correlations decay as $t_2$ increases and separates the $t_1$ and $t_3$ regions. Note that, as discussed in the text, 3-time correlations at $t_2=0$ are physically 2-time measurements of a modified operator, and could have been obtained without resorting to a 3-time simulation.}
\vspace{-6pt}
\end{figure*}

Already this simple spin-boson illustration allows us to make two important observations about kernel lifetimes and the relative sizes of 1-, 2-, and 3-time kernels. First, the $t_2$ slices in Fig.~\ref{fig:3timeprop}, right, display a decreasing lifetime as $t_2$ increases: as a generalization of the 1-time cutoff, $\tau_\mc{K}$, the total magnitude of the multitime kernel decreases to zero with increasing $t_1+t_2+t_3$, as is evident in the reducing area of the $t_1+t_3$ `triangle' down the column. We empirically observe the relationship $\tau_{1,\rmm{max}} + \tau_{2,\rmm{max}} + \tau_{3,\rmm{max}} \approx \tau_\mc{K}$. This means only $\sfrac{1}{4}$ of the total $\bm{\mc{C}}^{B_1,B_2}(t_1,t_2,t_3)$ `cube' needs to be simulated to obtain the full 3-time kernel. Second, we observe an inequality where $\bm{\mc{K}}(t) > \bm{\mc{K}}^B(t_2, t_1) > \bm{\mc{K}}^{B_2, B_1}(t_3, t_2, t_1)$. If one takes a full-index sum over the absolute value of the 1-, 2-, and 3-time contributions to the propagator, one obtains 100\%, 14\%, and 5\% of the same sum over the full propagator itself. Further memory effects are of decreasing importance as one might have guessed, and the fact that the sum comes to over 100\% implies there is significant cancellation between terms.

\vspace{-12pt}
\subsection{EET Dimer 2D-spectra}\label{sec:eet_results}
\vspace{-6pt}

We now turn to a system where obtaining the continuous memory kernels is impossible owing to an unaffordable growth in space complexity to store. Like the spin-boson model, the EET dimer is a 2-site model. We include a zero-excitation ground state and the one- and two-exciton manifolds. This results in a $4\times 4$ reduced density matrix, and so a $16\times 16$ Liouville space for all the propagators and memory objects. Furthermore, the two parameter regimes we investigate require more time steps to reach Markovianity. Specifically, running a $1000$~fs 1D simulation with a $\delta t = 1$~fs time step reveals that, in the first example, $\tau_\mc{K}\simeq 528$~fs. This is $\sim 4\times$ more than the fully-continuous spin-boson example above, and the time complexity of the 3-time propagator scales as $\mathcal{O}(N^3)$. As such, this projection operator-based approach generates objects that are $4^5=1024$~times larger than those of our previous spin-boson model example. This exceeds our local memory resources by orders of magnitude. For reference, a $(376\times47\times376\times16\times16)\simeq 2^{31}$~element array demands 25.3 GB. Furthermore, the simulations required to extract the memory objects also scale (at minimum) as $\mc{O}(N^3)$ in time, and also with the size of the Liouville space (one factor of $4$). Most significant about this analysis is that this steep rise in computational cost arises in considering \textit{only} what is practically the next most affordable model. Indeed, the continuous time implementation of this multitime decomposition\cite{Ivanov2015a} did not include the two-exciton manifold, making it fractionally cheaper. Yet, chromophore aggregates in nature can have $10$-$100$ chromophores,\cite{Durrant1995, Jang2018b, Biswas2022a} and there is no guarantee they will reach Markovianity in $\mathcal{O}(10^3)$ time steps, although in some well-studied examples memory kernels can decay on the order of $10$--$100$~fs although kinetic exchange across bottleneck states and subsequent equilibration can take on the order of $10$~ps.\cite{Pfalzgraff2019a}

To address this, we leverage our discrete-time description to coarsen all time axes. We choose to make $\delta t_1 = \delta t_3 = 6\delta t$ and $\delta t_2 = 48\delta t$ as a first attempt. The savings arise from two sources. First, we have a $6\times6\times48=1728$-fold improvement in space, which makes the cost to store the multitime response of this system \textit{even lower} than that of the fully continuous spin-boson example. Second, while the HEOM time step cannot itself be increased due to the requirements of the integrator, there is no need to simulate time series with measurements at times that are not be used in the final construction, which means the simulation step enjoys a speed-up of $\delta t_2 \delta t_3 / \delta t^2 = 288$.

\begin{figure*}[!t]
\vspace{-10pt}
\begin{center} 
    \hspace{-20pt}
    \resizebox{.38\textwidth}{!}{\includegraphics{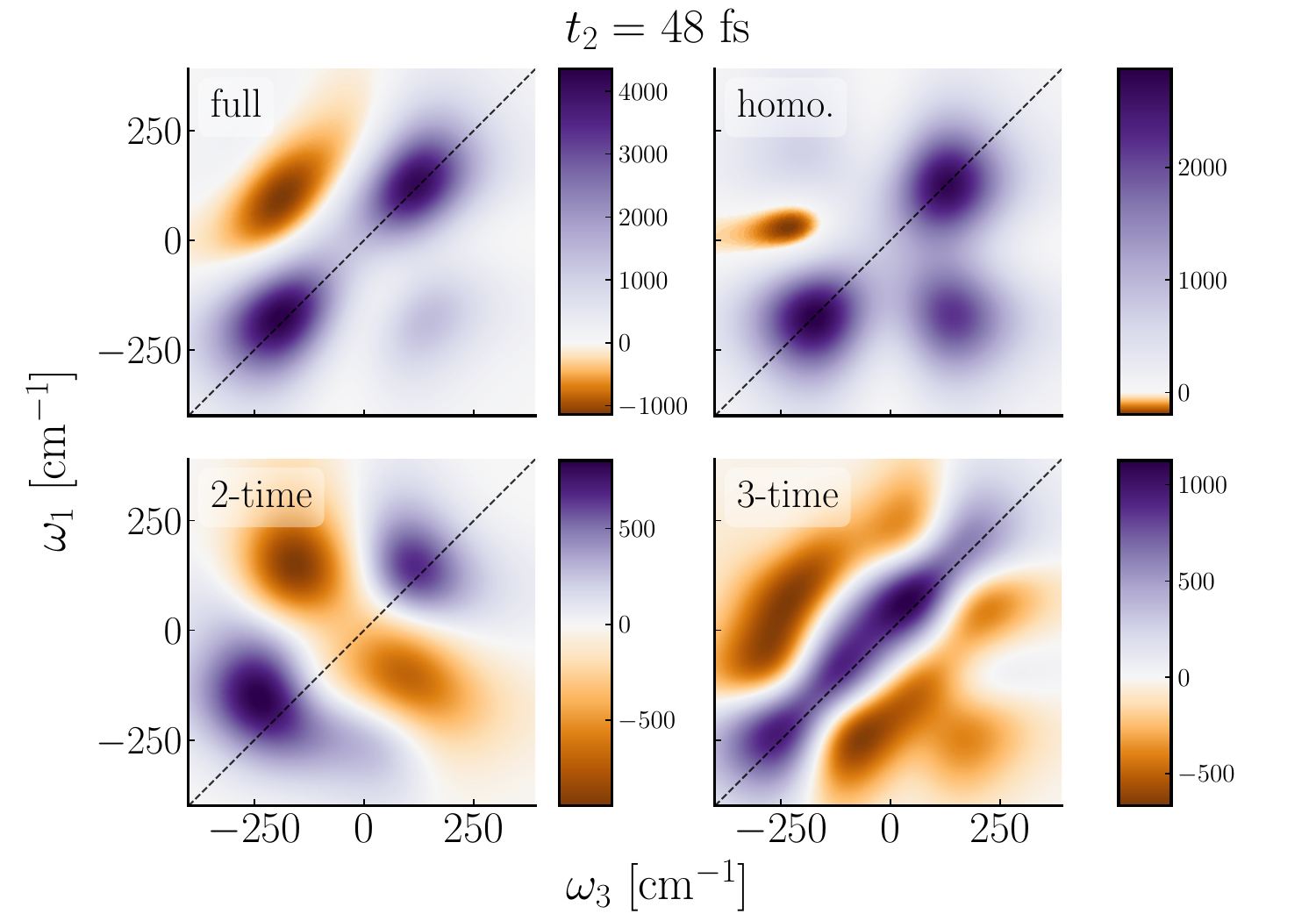}}\hspace{-9pt}\vline \hspace{+6pt}\resizebox{.33\textwidth}{!}{\includegraphics[trim={100pt 0 0 0}, clip]{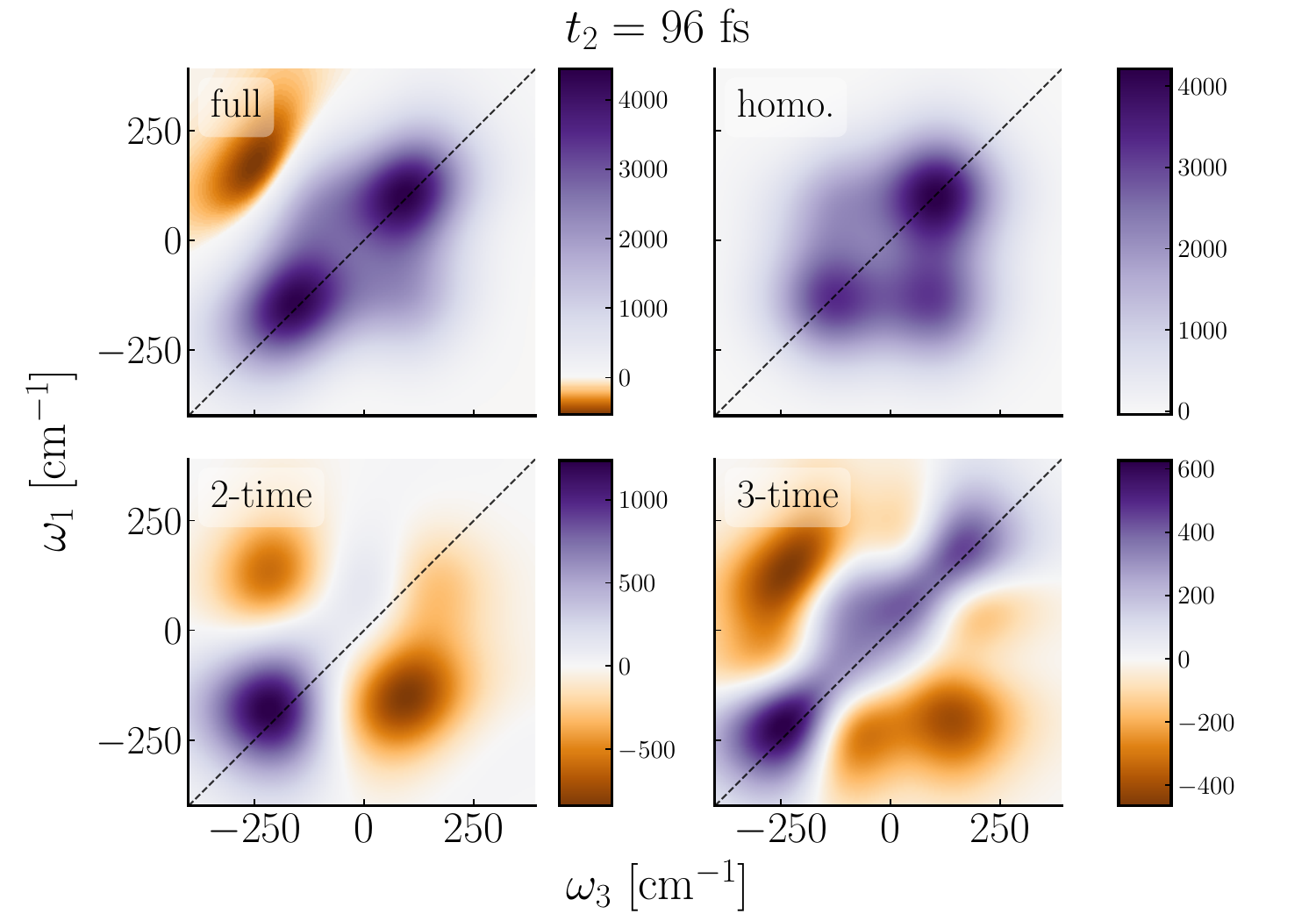}}\hspace{-9pt}\vline \hspace{+3pt}
    \resizebox{.33\textwidth}{!}{\includegraphics[trim={100pt 0 0 0}, clip]{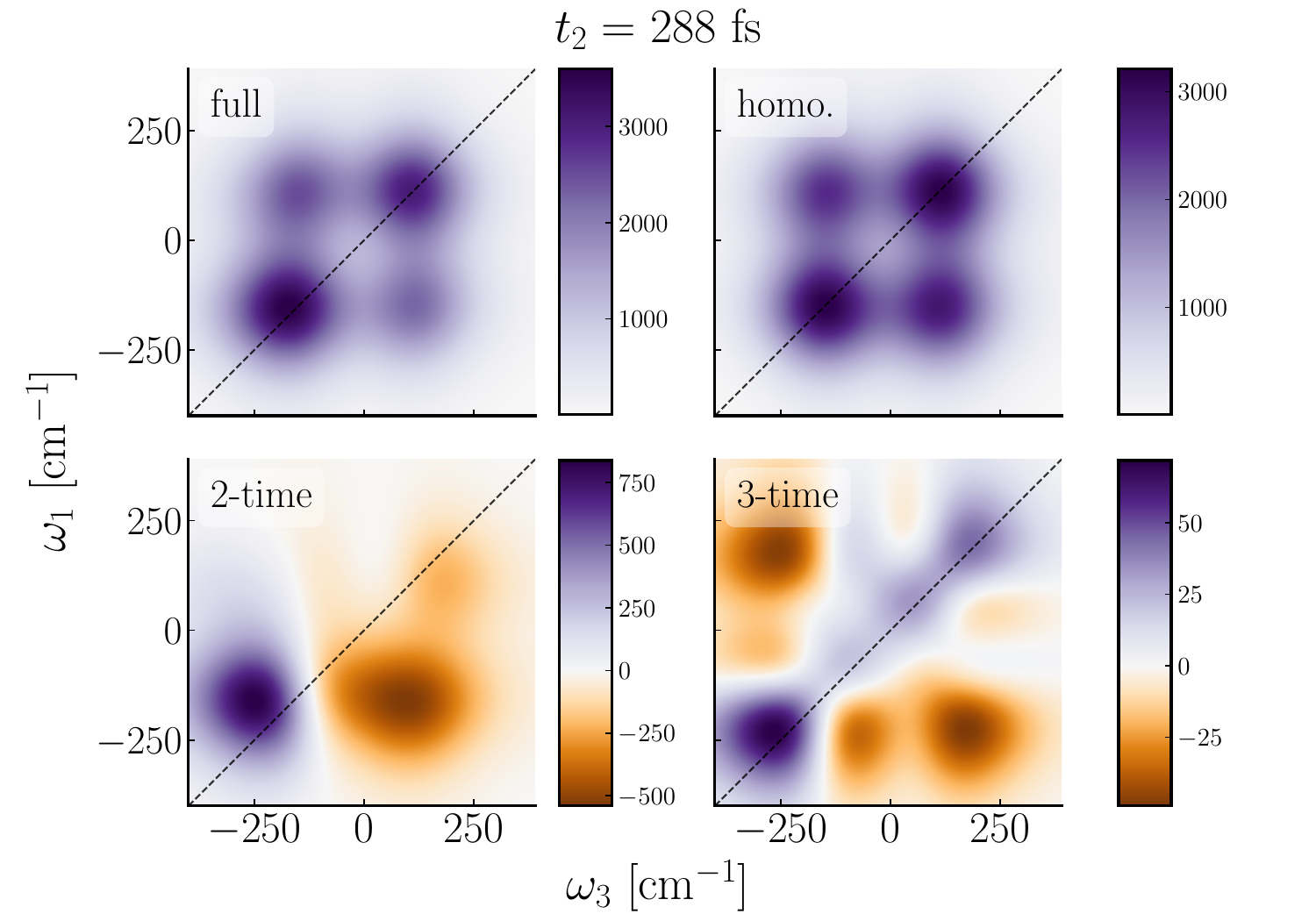}}
\end{center}
\vspace{-16pt}
\caption{\label{fig:damped} HEOM 2D spectra, including decomposition into 1-, 2-, and 3-time correlations, at different $t_2$ wait times for the EET dimer with $\pm\epsilon = \pm50~\rmm{cm}^{-1}$ and $\Delta = 100~\rmm{cm}^{-1}$, while $\lambda = 50~\rmm{cm}^{-1}$; $\tau = 100~\mathrm{fs}$ and $T=298$~K, $K=0$ and $L=9$. \textbf{Left}: Very early-time spectra, $t_2=48$~fs, showing a strong on-diagonal absorption for the full spectrum with a smaller, but not insignificant, off-diagonal absorption arising from 2- and 3-time correlations. \textbf{Middle}: Early-time spectra, $t_2=96$~fs, where off-diagonal absorption grows in. 2-time correlations are quantitatively asymmetric for the off-diagonal signals. \textbf{Right}: Later-time spectra, $t_2=288$~fs, where the 3-time correlations have insignificant magnitude and the cross peaks are fully visible.}
\vspace{-6pt}
\end{figure*}

Under the linear response formulation of spectroscopy,\cite{Mukamel1999a, Cho2009} the 2D optical response is the sum of two terms, $I=I_- + I_+$, whose forms are given by
\begin{equation}\label{eq:Iphasing}
    I_\mp(\omega_1, t_2, \omega_3) = \mathrm{Re}\iint_0^\infty \dd{t_1}\,\dd{t_3} \e{i(\omega_3t_3\mp\omega_1t_1)} R_\mp(t_3, t_2, t_1),
\end{equation}
where
\begin{equation}\label{eq:rephase}
    R_- (t_3, t_2, t_1) = \mathrm{tr_s}\left\lbrace \mu_\rmm{L} \bm{\mc{C}}^{\mu_\rmm{R}^\times, \mu_\rmm{R}^\times}_{t_3, t_2, t_1} \mu_\rmm{L}^\times |0\rangle\langle0|  \right\rbrace
\end{equation}
is the rephasing contribution to the spectrum and
\begin{equation}\label{eq:nonrephase}
    R_+ (t_3, t_2, t_1) = \mathrm{tr_s}\left\lbrace \mu_\rmm{L} \bm{\mc{C}}^{\mu_\rmm{R}^\times, \mu_\rmm{L}^\times}_{t_3, t_2, t_1} \mu_\rmm{R}^\times |0\rangle\langle0|  \right\rbrace
\end{equation}
is the non-rephasing contribution. The Hamiltonian is excitation-conserving, so only light pulses can move population between manifolds. The transition dipole for the two excitons are different, given by $[1.0, -0.2]$. While the model does not account for spontaneous emission, it captures nonadiabatic energy transfer among chromophores.

\vspace{-14pt}
\subsubsection{Weak Coupling Regime}
\vspace{-8pt}
We begin with a very weakly coupled system where the reorganization energy of the baths is only $2$~cm$^{-1}$. Although this example is in an unrealistically small coupling regime, it is nonetheless instructive. In the $500$~fs of data for which we simulate, $R_-$ completes fewer than $2$ full oscillations which are only weakly damped (see Fig.~\ref{fig:underdamped}, left). Therefore, extending the upper limit on the integrals of Eq.~\ref{eq:Iphasing} to infinity is not possible. In other words, the spectrum from these data (when zero-padded) suffers from artifacts, appearing blurry. In this underdamped regime, one has to extend the simulations by around $1700$~fs to reach equilibrium, which means use of the MNZ equation results in a savings factor of just under $5$, on both $t_1$ \textit{and} $t_2$ axes. Additionally, we can decompose the resulting spectrum into the contributions arising from 1-, 2-, and 3-time correlations, which we display in Fig.~\ref{fig:underdamped}, middle. In this coupling regime, almost the entire spectrum is composed of 1-time quantities. Even at time zero, higher-order contributions amount to less than a sixth of the total peak height; at time zero, all higher-order contributions can be obtained from 2-time simulations. Later in $t_2$, the 3-time contributions decay fairly rapidly to zero, as shown in Fig.~\ref{fig:underdamped}, right. What is more, 3-time contributions have a nearly indistinguishable structure from 1-time contributions, which further supports the conclusion that they are unnecessary to describing these 2D spectra. Ultimately, these spectra lack structure along $t_2$ and are highly symmetric as this example lies in the weak coupling extremum of the parameter regime and there is only minor dissipation. 

\vspace{-14pt}
\subsubsection{Intermediate Coupling Regime}
\vspace{-8pt}
We now turn to a more experimentally reasonable value for the reorganization energy, $\lambda = 50~\rmm{cm}^{-1}$, which has resulted in more structured, informative 2D signals in previous studies.\cite{Chen2010h, Fetherolf2017b} A large reorganization energy means that nuclei can rearrange to offer a greater stabilization to the excitation, leading to its localization. In comparison to the previous case, the peaks become broad, representing the participation of many oscillation frequencies in $R_\mp$. Since the transition dipole is not the same for both excitons, the initial population after the first light pulse is not equal, resulting in an unequal signal at early $t_2$ times. As $t_2$ increases, the population is given time to flow between the excitons, leading to both the appearance of the cross peaks in the spectrum and an oscillating intensity of the diagonal peaks as the system and bath equilibrate. This progression can be observed in the top-left panels of Fig.~\ref{fig:damped}. 

In this regime of stronger coupling, $\tau_\mc{K}$ is shorter, only around $300$~fs. With the dynamics being much more strongly damped, extension of $R_\mp$ in $t_1$ and $t_3$ is unnecessary to obtain clear spectra, which is why this parameter regime was affordable in previous works with fixed choices of $t_2$. However, we can go further than previous work by decomposing the spectra into 1-, 2-, and 3-time correlations. Since the details of these multitime correlations dictate how the system and the bath equilibrate, one should expect to see much richer information in this regime. Indeed, one can observe from the progression of the upper-right panels in Fig.~\ref{fig:damped} that the \textit{1-time contributions become purely absorptive}, and increasingly more 4-fold symmetrical across the $t_2$ range, in contrast to the full spectra. This means that 1-time contributions suggest that the transfer rates between the excitons are the same, which clearly violates detailed balance. Physically we can explain this as being due to the ignorance of the baths' equilibration to the $t_1$ region before entering the $t_2$ region, and similarly for $t_2$ and $t_3$. This is precisely the meaning of the inhomogeneous terms arising from interactions with the light pulses. Moreover, we see from the lower panels that the role of the 2- and 3-time correlations is exactly to suppress the off-diagonal peaks while reinforcing the diagonal ones. However, it is only the 2-time correlations that differentiate between the high and low energy excitons, as 3-time correlations do not qualitatively break the symmetry of the off-diagonal peaks' intensities. In addition, the 3-time correlations decay with time, as we previously observed, while the 2-time contributions have a relatively stable intensity (like the 1-time contributions). Consistent with the observations of Sec.~\ref{ssec:spin-boson_results_3time}, the 3-time information (the lower-right panels) can be ignored for $t_2 \gtrapprox \tau_\mc{K} \simeq 300~\rmm{fs}$. 

\begin{figure}[!t]
\vspace{-12pt}
\begin{center} 
    \resizebox{.38\textwidth}{!}{\includegraphics{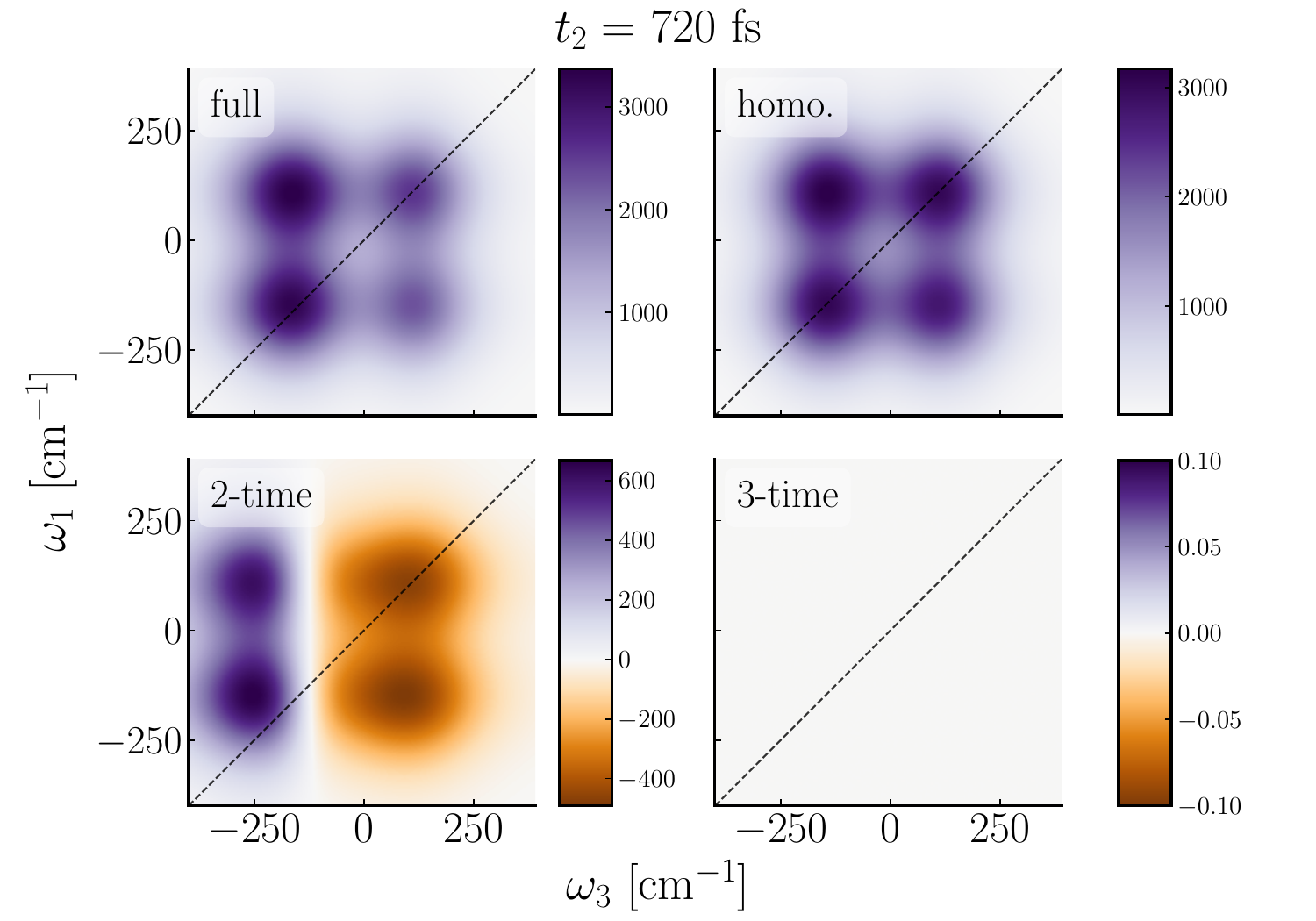}}
\vspace{-6pt}
\caption{\label{fig:extend_t2} GQME 2D spectrum extending the data of Fig.~\ref{fig:damped} along the $t_2$ axis. Since $t_2 \gg \tau_\mc{K}$, the 3-time correlations are set to zero. We observe the expected column-like spectrum at the lower energy resonance as the system undergoes long-time system-bath equilibration. The peaks' shift is encoded in the 2-time correlations, as the homogeneous term is qualitatively the same as it was at $t_2 = 288$~fs.}
\end{center}
\vspace{-18pt}
\end{figure}

The vanishing 3-time contribution also has practical implications.\cite{Ivanov2015a} Consider predicting the spectra at very long $t_2$, where one would expect the majority of the signal to come from the lower energy exciton (a column-like spectrum). Since the 3-time correlations have already become insignificant on the previous simulated timescale, extension of these response functions in $t_2$ only requires 2-time quantities. We display the results of extending the $t_2$ axis to 720~fs, around two-and-a-half times longer than we simulated, in Fig.~\ref{fig:extend_t2}. We can also directly simulate the result at this particular $t_2$ time (i.e., a single 3-time simulation where $\delta t_2 = 720$~fs) and obtain excellent quantitative agreement with the neglected 3-time terms being less than $\sim 0.5$ units in magnitude, compared to $\sim 3000$ for the total signal. In addition to observing a column-like structure, we can also see that the $\omega_3$ redshift of the spectrum arises from the 2-time correlations, as the 1-time term has no knowledge of the relaxation of the bath in the presence of only one of the excitons. Finally, we again observe that $\tau_{1,\rmm{max}} + \tau_{2,\rmm{max}} + \tau_{3,\rmm{max}} \approx \tau_\mc{K}$, in agreement with the above observation that  $\tau_{2,\rmm{max}} \leq \tau_\mc{K}$. In other words, simulation of the 3-time propagator up to $\tau_\mc{K}$ appears to be always sufficient to extend the spectra along any and all axes.

\vspace{-12pt}
\section{Conclusion}
\vspace{-6pt}

We have explained why a continuous-time projection operator-based decomposition of multitime correlation functions offers benefits, such as interpretability and the advantageous property that high-order memory kernels decay on timescales at worst as long as the 1-time kernels. Yet these are computationally demanding to produce as they require sufficiently fine temporal resolution across all time axes to enable the accurate extraction of the kernels from the time derivatives of the multime correlation functions. We also saw that a discrete implementation naturally arises when employing the TTM, which removes discretization error, potentially leading to an efficient implementation of these multitime kernel ideas. However, the formulation of the TTM as an iterative correction to otherwise Markovian evolution folds into the transfer tensor inhomogeneous terms that naturally arise from a projection operator-based decomposition of the multitime correlation functions, obfuscating their meaning and entangling their lifetimes. Instead, our work combines the strengths of both approaches, leading to a highly efficient discretization of the projection operator-based decomposition that exploits the interval size-independent discrete component decomposition of convolution integrals, offering a means to develop an iterative extraction of the underlying kernels without the need for derivatives or TTM-like appeals to corrections to the deviation from Markovianity. Indeed, our approach can be understood as a discrete representation of the deviation away from the Gaussian decomposability of multitime correlations within a subspace spanned by an MNZ-like projection operator. This ultimately leads to an \textit{exact} decomposition of multitime correlations across interaction times. 

We have demonstrated that this approach exactly recovers benchmark results for the SB and EET models and has the potential to greatly reduce the computational cost of calculating multitime correlation functions at the heart of multidimensional spectroscopies. What is more, this decomposition enables one to pinpoint the precise windows of time and correlations that govern the system's evolution and eventual relaxation.

The high cost of predicting multitime correlation functions is a general feature of just about all dynamical methods, and not just HEOM. For example, \textit{all} multitime methods scale with the resolution of each time axis. While our projection operator-based method can only alter the $\mathcal{O}(t_{1,\rmm{max}}t_{2,\rmm{max}}t_{3,\rmm{max}})$ scaling after some cutoff based on $\tau_\mc{K}$, it can address the prefactor which can itself scale nonlinearly in other Hamiltonian parameters and the time step employed along each time axis. Taking $\delta t_2$ an order of magnitude coarser than $\delta t_1$,\cite{Son2017} which itself is coarsened by $10\times$ with respect to the simulation time step, the cost of this method is reduced by $\mathcal{O}(10^3)$ when compared to the full-resolution, continuous-time implementation.\cite{Ivanov2015a} Compared to a direct simulation, the saving is most evident for long $t_2$ times, where the GQME method only requires 2-time quantities calculated up to $\tau_\mc{K}$---never requiring a 3-time simulation. Even the computation a single long-$t_2$ spectrum requires that every $t_1$ time have a much longer $t_2$ simulation spawn off of it, which will prove prohibitive for non-linear scaling methods, and deleterious for trajectory-based methods like surface hopping and Ehrenfest dynamics.\cite{Hanna2009a, McRobbie2009a, VanDerVegte2013a, Tempelaar2013a, Provazza2018b, Polley2021a, Mannouch2022, Loring2022, Kim2022a, Atsango2023, Mondal2023} Hence, the combination of these methods with the present approach offers exciting possibilities.

Our results also pose interesting questions for the analysis of 2D spectra. After all, an important reason to build a 2D spectrometer is to capture the structure and dynamics of materials via 3-time correlators. Yet, we have shown that these are only briefly relevant, and could simplified to 2-time correlations for a number of empirical questions. Our results thus have the potential to refocus the multidimensional question on such 2-time contributions and the physics and chemistry they encode. 

\vspace{-12pt}
\begin{acknowledgments}
\vspace{-8pt}
This work was partially supported by an Early Career Award in CPIMS program in the Chemical Sciences, Geosciences, and Biosciences Division of the Office of Basic Energy Sciences of the U.S.~Department of Energy under Award DE-SC0024154. We also acknowledge start-up funds from the University of Colorado Boulder.
\end{acknowledgments}

\pagebreak
\appendix
\section{Unravelling of the 2-~and 3-time Memory Kernels}\label{app:auxiliary_forms}
\vspace{-8pt}

Here, we provide derivations that unfold the continuous-time memory kernels derived in Ref.~\onlinecite{Ivanov2015a} for the 2- and 3-time correlation functions. Unlike Ref.~\onlinecite{Ivanov2015a}, we employ the transparent notation for a generalized Mori-type projector $\mathcal{P} = | \mathbf{A})( \mathbf{A}|$ adopted in Refs.~\onlinecite{Montoya-Castillo2016, Kelly2016}. This ensures that our formulation is general and independent of the specific choice of projection operator.

\vspace{-10pt}
\subsection{Auxiliary form of 2-time kernel}
\vspace{-6pt}
In the main text we arrived at a projection operator form for the 2-time memory kernel
\begin{equation}\tag{\ref{eq:2timemem}}
    \bm{\mathcal{K}}^B_{t_2,t_1} \equiv (\mathbf{A}|\mc{L} \e{\mc{Q}\mc{L}t_2} \mc{Q} B \e{\mc{Q}\mc{L}t_1} \mc{Q}\mc{L} |\mathbf{A}).
\end{equation}

We can self-consistently expand this memory kernel by invoking the Dyson identity, as has been done in the context of 1-time memory kernels,\cite{Shi2003a, Montoya-Castillo2016}
\begin{align}\label{eq:dyson_identity}
    \e{\mc{Q}\mc{L}\tau} &= \e{\mc{L}\tau} - \int_0^\tau \dd{s} \e{\mc{L}s} \mc{P}\mc{L} \e{\mc{Q}\mc{L}(\tau-s)}  \\
            &= \e{\mc{L}\tau} - \int_0^\tau \dd{s} \e{\mc{QL}s} \mc{P}\mc{L} \e{\mc{L}(\tau-s)},
\end{align}
where the $s$ and $\tau-s$ placement can also be readily exchanged by substitution. The idea is to substitute the appropriate expansion such that the original expression appears nested within the second resulting term. At this point, one can use the existing machinery\cite{Pfalzgraff2019a} to obtain $\bm{\mc{K}}^B_{t_2,t_1}$ by matrix inversion. In the paper, we use the fully discrete TTM instead. One also needs to place the complementary projector $\mc{Q}$ in the correct location before the expansion (it does not commute with $\e{\mc{L}t}$). For this, we note that
\begin{equation}
    \e{\mc{QL}t}\mc{Q} = \mc{Q}\e{\mc{LQ}t}.
\end{equation}
This is proved by Taylor expansion. Further, since $\mc{Q}^2 = \mc{Q}$ by idempotency of projectors, $\mc{Q}\e{\mc{QL}t}\mc{Q} = \mc{Q}\e{\mc{LQ}t}\mc{Q}$.

We perform the expansion on the left (label $L$) propagator first without loss of generality,
\begin{align}
    \bm{\mathcal{K}}^B_{t_2,t_1} &\equiv (\mathbf{A}|\mc{L} \e{\mc{Q}\mc{L}t_2} \mc{Q} B \e{\mc{Q}\mc{L}t_1} \mc{Q}\mc{L} |\mathbf{A}) \tag{\ref{eq:2timemem}}\\
    &= (\mathbf{A}|\mc{L} \e{\mc{L}t_2} \mc{Q} B \e{\mc{Q}\mc{L}t_1} \mc{Q}\mc{L} |\mathbf{A}) \nonumber\\
    &~~~~~- \int_0^{t_2}\ris\dd{s}\,(\bm{A}|\mc{L} \e{\mc{L}s} \mc{PL} \e{\mc{QL}(t_2-s)} \mc{Q} B \e{\mc{Q}\mc{L}t_1} \mc{Q}\mc{L} |\bm{A}) \label{eq:k3line} &\\
    &\equiv \bm{\mathcal{K}}^B_{t_2,t_1;L} - \int_0^{t_2}\ris\dd{s}\, \dot{\bm{\mc{C}}}(s) \bm{\mathcal{K}}^B_{t_2-s,t_1}, \label{eq:2tmem,nest1}
\end{align}
where we have identified $\dot{\bm{\mc{C}}}$ and found the nested form of Eq.~\ref{eq:2timemem} to its right. This is $\dot{\bm{\mc{C}}} = \frac{\rmm{d}}{\rmm{d}t}(\mathbf{A}|e^{\mathcal{L}t}|\mathbf{A})$, obtained from the simulation of the 1-time correlation function defined by the projection operator, $\mc{P}$. Note how this term does not reference the measurement, despite contributing to the evolution \textit{after} the measurement. This is the key feature of the method.

Since $\bm{\mc{K}}^B_{t_2,t_1;L}$ still contains one projected propagator, we apply the Dyson identity again to obtain
\begin{align}
    \bm{\mc{K}}^B_{t_2,t_1;L} &\equiv (\mathbf{A}|\mc{L} \e{\mc{L}t_2} \mc{Q} B \e{\mc{Q}\mc{L}t_1} \mc{Q}\mc{L} |\mathbf{A}) \label{eq:2tmemL}\\
    &= (\mathbf{A}|\mc{L} \e{\mc{L}t_2} \mc{Q} B \mc{Q} \e{\mc{LQ}t_1} \mc{L} |\mathbf{A}) \\
    &= (\mathbf{A}|\mc{L} \e{\mc{L}t_2} \mc{Q} B \mc{Q} \e{\mc{L}t_1} \mc{L} |\mathbf{A})\nonumber\\
    &~~~~~- \int_0^{t_1}\ris\dd{s}\,(\mathbf{A}|\mc{L} \e{\mc{L}t_2} \mc{Q} B \mc{Q} \e{\mc{LQ}(t_1-s)} \mc{LP} \e{\mc{L}s} \mc{L} |\mathbf{A}) \\
    &\equiv \bm{\mc{K}}^B_{t_2,t_1;LR} - \int_0^{t_1}\ris\dd{s}\, \bm{\mc{K}}^B_{t_2,t_1-s;L} \dot{\bm{\mc{C}}}(s).  \label{eq:2tmem,nest2}
\end{align}
The projection-free term is this `left-right' expanded object. To assess this we must make the $\mc{Q} = \mathbb{1} - \mc{P}$ substitution twice, arriving at a four-term expression,
\begin{align}
    \bm{\mc{K}}^B_{t_2,t_1;LR} &\equiv (\mathbf{A}|\mc{L} \e{\mc{L}t_2} \mc{Q} B \mc{Q} \e{\mc{L}t_1} \mc{L} |\mathbf{A}) \label{eq:2tmemLR}\\
        &= (\mathbf{A}|\mc{L} \e{\mc{L}t_2} B \e{\mc{L}t_1} \mc{L} |\mathbf{A})\nonumber\\
        &~~~~~- (\mathbf{A}|\mc{L} \e{\mc{L}t_2} \mc{P}B \e{\mc{L}t_1} \mc{L} |\mathbf{A}) \nonumber\\
        &~~~~~- (\mathbf{A}|\mc{L} \e{\mc{L}t_2} B \mc{P} \e{\mc{L}t_1} \mc{L} |\mathbf{A}) \nonumber\\
        &~~~~~+ (\mathbf{A}|\mc{L} \e{\mc{L}t_2} \mc{P} B \mc{P} \e{\mc{L}t_1} \mc{L} |\mathbf{A}) \\
        &= \partial_{t_1}\partial_{t_2}\bm{\mc{C}}^B_{t_2,t_1} - \dot{\bm{\mc{C}}}(t_2){}^B\dot{\bm{\mc{C}}}(t_1) \nonumber\\
        &~~~~~- \dot{\bm{\mc{C}}}^B(t_2)\dot{\bm{\mc{C}}}(t_1) + \dot{\bm{\mc{C}}}(t_2)\bm{\mc{C}}^B(0)\dot{\bm{\mc{C}}}(t_1), \label{eq:KLR}
\end{align}
where the equation defines this ${}^B\bm{\mc{C}}$ versus $\bm{\mc{C}}^B$ notation, such that the latter is the $\bm{\mc{C}}$ matrix when $B$ acts at the beginning of the simulation, at $t_1=0$. This is analogous to the 1-time problem with $B\mc{R}$ on the right. Only the first term requires a 2-time simulation as it is the double (mixed) derivative of the 2-time propagator.

The procedure is then:
\begin{itemize}
    \item Get the first term in Eq.~\ref{eq:2timeprop_decomp} from 1-time quantities within the simulation-derived $\bm{\mc{C}}^B_{t_1,t_2}$.
    \item Get $\bm{\mc{K}}^B_{t_2,t_1;LR}$ from operating on $\bm{\mc{C}}^B_{t_1,t_2}$ and the 1-time quantities with derivatives according to Eq.~\ref{eq:KLR}.
    \item Perform the inversion Eq.~\ref{eq:2tmem,nest2} on the $t_1$ index to get $\bm{\mc{K}}^B_{t_2,t_1;L}$. This does it for all the $t_2$ values at once.
    \item Perform the inversion Eq.~\ref{eq:2tmem,nest1} on the $t_2$ index to get $\bm{\mc{K}}^B_{t_2,t_1}$. This does it for all the $t_1$ values at once.
\end{itemize}
Assuming enough time is recorded for the 2-time kernel to decay to zero, one can perform the double convolution integral to get the second term in Eq.~\ref{eq:2timeprop_decomp}, and therefore the full propagator, for any later time.

\pagebreak
\subsection{3-time Memory Kernel}

Beyond the objects introduced for the 2-time correlation function in Appendix~\ref{app:auxiliary_forms}, the 3-time correlation function in Eq.~\ref{eq:3timeprop},
\begin{equation}
    \bm{\mc{C}}_{t_3, t_2, t_1}^{B_2,B_1} = (\mathbf{A}| \e{\mc{L}t_3} B_2 \e{\mc{L}t_2} B_1 \e{\mc{L}t_1} |\mathbf{A}), \label{eq:3U}
\end{equation}
requires the construction of the 3-time memory kennel,
\begin{equation}
    \bm{\mc{K}}_{t_3, t_2, t_1}^{B_2, B_1} = (\mathbf{A}| \mc{L}\e{\mc{QL}t_3} \mc{Q} B_2 \e{\mc{QL}t_2} \mc{Q} B_1 \e{\mc{QL}t_1} \mc{Q} \mc{L} |\mathbf{A}). \tag{\ref{eq:3timemem}}
\end{equation}

We condense this expression by identifying the 2-time objects of Eq.~\ref{eq:2timeprop_decomp}, yielding
\begin{align}
    \bm{\mc{C}}_{t_3, t_2, t_1}^{B_2,B_1} &= \bm{\mc{C}}(t_3) \bm{\mc{B}}_2 \bm{\mc{C}}^{B_1}_{t_2, t_1} + \bm{\mc{C}}(t_3-\tau_3)\bm{\mc{K}}^{B_2}_{\tau_3, \tau_2}\bm{\mc{C}}^{B_1}_{t_2-\tau_2, t_1} \nonumber\\
                                    &~~~ + \bm{\mc{C}}(t_3-\tau_3) \bm{\mc{K}}_{\tau_3, t_2, \tau_1}^{B_2, B_1} \bm{\mc{C}}(t_1-\tau_1), \label{eq:U3_condensed}
\end{align}
where the homogeneous term is no longer present and the first two terms incorporate both 1-time and 2-time contributions. The final term, containing the 3-time contribution, is the same in both expressions.

The treatment in the previous section must now be repeated for all three projected propagators. We begin again on the left side,

\begin{widetext}
\begin{align}
    \bm{\mc{K}}_{t_3, t_2, t_1}^{B_2, B_1} &= (\mathbf{A}| \mc{L} \e{\mc{L}t_3} \mc{Q} B_2 \e{\mc{QL}t_2} \mc{Q} B_1 \e{\mc{QL}t_1} \mc{QL} |\mathbf{A}) -\int_0^{t_3}\ris\dd{s_3} (\mathbf{A}|\mc{L}\e{\mc{L}s_3} \mc{PL} \e{\mc{QL}(t_3-s_3)} \mc{Q} B_2 \e{\mc{QL}t_2} \mc{Q} B_1 \e{\mc{QL}t_1} \mc{QL}|\mathbf{A}) \\
        &\equiv \bm{\mc{K}}_{t_3, t_2, t_1;L}^{B_2, B_1} - \int_0^{t_3}\ris\dd{s_3} \dot{\bm{\mc{C}}}(s_3) \bm{\mc{K}}_{t_3-s_3, t_2, t_1}^{B_2, B_1}.
\end{align}
We now focus on the left-expanded kernel, which becomes 
\begin{align}
    \bm{\mc{K}}_{t_3, t_2, t_1;L}^{B_2, B_1} &= (\mathbf{A}| \mc{L} \e{\mc{L}t_3} \mc{Q} B_2 \e{\mc{L}t_2} \mc{Q} B_1 \e{\mc{QL}t_1} \mc{QL} |\mathbf{A}) -\int_0^{t_2}\ris\dd{s_2} (\mathbf{A}|\mc{L} \e{\mc{L}t_3} \mc{Q} B_2 \e{\mc{L}s_2}\mc{PL}\e{\mc{QL}(t_2-s_2)} \mc{Q} B_1 \e{\mc{QL}t_1} \mc{QL}|\mathbf{A}) \\
        &\equiv \bm{\mc{K}}_{t_3, t_2, t_1;LM}^{B_2, B_1} - \int_0^{t_2}\ris\dd{s_2} \left(\partial_{t_3} \bm{\mc{C}}_{t_3,s_2}^{B_2} - \dot{\bm{\mc{C}}}(t_3){}^{B_2}\bm{\mc{C}}(s_2)\right) \bm{\mc{K}}^{B_1}_{t_2-s_2,t_1}. \label{eq:3tsingle_conv}
\end{align}
This expression has new (single) derivatives of some of these 2-time quantities. Other than these details, the structure of these terms is reminiscent of that seen in the 2-time problem. The final auxiliary memory term is similar to the second inversion step of the 2-time problem, where we commute the $\mc{Q}$ to flip the propagator appropriately,
\begin{align}
    \bm{\mc{K}}_{t_3, t_2, t_1;LM}^{B_2, B_1} &= (\mathbf{A}|\mc{L} \e{\mc{L}t_3} \mc{Q} B_2 \e{\mc{L}t_2} \mc{Q} B_1 \mc{Q} \e{\mc{LQ}t_1} \mc{L}|\mathbf{A}) \\
                &= (\mathbf{A}|\mc{L} \e{\mc{L}t_3} \mc{Q} B_2 \e{\mc{L}t_2} \mc{Q} B_1 \mc{Q} \e{\mc{L}t_1} \mc{L}|\mathbf{A}) -\int_0^{t_1}\ris\dd{s_1}(\mathbf{A}|\mc{L} \e{\mc{L}t_3} \mc{Q} B_2 \e{\mc{L}t_2} \mc{Q} B_1 \mc{Q} \e{\mc{LQ}s_1} \mc{LP} \e{\mc{L}(t_1-s_1)}  \mc{L}|\mathbf{A}) \\
                &\equiv \bm{\mc{K}}_{t_3, t_2, t_1;LMR}^{B_2, B_1} - \int_0^{t_1}\ris\dd{s_1} \bm{\mc{K}}_{t_3, t_2, s_1;LM}^{B_2, B_1} \dot{\bm{\mc{C}}}(t_1-s_1).
\end{align}
The final step is to make substitutions for all the remaining instances of $\mc{Q}$, which will yield $8$~terms; these terms are
\begin{align}
    \bm{\mc{K}}_{t_3, t_2, t_1;LMR}^{B_2, B_1} &= \partial_{t_3} \partial_{t_1}\bm{\mc{C}}_{t_3, t_2, t_1}^{B_2,B_1} -\left( 
        \dot{\bm{\mc{C}}}(t_3) \partial_{t_1}{}^{B_2}\bm{\mc{C}}^{B_1}_{t_2,t_1} + 
        \partial_{t_3}\bm{\mc{C}}^{B_2}_{t_3,t_2}  {}^{B_1}\dot{\bm{\mc{C}}}(t_1) +
        \partial_{t_3}\bm{\mc{C}}^{B_2|B_1}_{t_3,t_2}\dot{\bm{\mc{C}}}(t_1)
        \right)  - \dot{\bm{\mc{C}}}(t_3) {}^{B_2}\bm{\mc{C}}(t_2) \bm{\mc{C}}^{B_1}(0) \dot{\bm{\mc{C}}}(t_1)  \nonumber\\
        &~~~ +\left(
            \dot{\bm{\mc{C}}}(t_3) {}^{B_2}\bm{\mc{C}}(t_2) {}^{B_1}\dot{\bm{\mc{C}}}(t_1) +
            \dot{\bm{\mc{C}}}(t_3) {}^{B_2}\bm{\mc{C}}^{B_1}(t_2) \dot{\bm{\mc{C}}}(t_1) +
            \partial_{t_3} \bm{\mc{C}}^{B_2}_{t_3,t_2} \bm{\mc{C}}^{B_1}(0) \dot{\bm{\mc{C}}}(t_1)
        \right),
\end{align}
where $\bm{\mc{C}}^{B_2|B_1}_{t_3,t_2}$ is the 2-time propagator starting from the $t_1=0$ instant $B_1$ measurement, which is given by $\bm{\mc{C}}^{B_2,B_1}_{t_3,t_2,0}$. The first term, which is the double derivative of the full propagator, is the term that must be extracted to the 3-time simulation.
\end{widetext}

\section{Computational Details}\label{app:comp}

We performed all HEOM calculations using the open-source pyrho code \cite{Berkelbach2020}. For the systems in the main text, the parameter regimes are, in order of appearance:
\vfill
\pagebreak

\begin{enumerate}
    \item Spin-boson models of Figs.~\ref{fig:2timeprop} and~\ref{fig:3timeprop}: $\epsilon = 1.0$, $\Delta = 1.0$, $k_\rmm{B}T = 10 \Delta$, $\omega_c = 2 \Delta$, $\lambda = 0.1\Delta$; in energy units where $\Delta = 1$. For the HEOM dynamics, we employed $K=0$ and $L=3$ with $\delta t=0.01$. For the dissipative bath, we used the Debye spectral density, i.e., $J(\omega) = 2\lambda\omega\omega_c/(\omega^2 + \omega_c^2)$. $B_1 = \sigma_x$ and $B_2 = \sigma_y$, the Pauli spin matrices.
        
    \item Excitation energy transfer dimer of Fig.~\ref{fig:underdamped}: $\epsilon = 50~\rmm{cm}^{-1}$, $\Delta = 100~\rmm{cm}^{-1}$, $T = 298~\rmm{K}$ ($\rmm{k}_\rmm{B} = 0.69352~\rmm{cm}^{-1}\rmm{K}^{-1}$ and $\hbar = 5308.8~\rmm{cm}^{-1}\rmm{fs}$), $\lambda=2$~cm$^{-1}$, $\tau=1/\omega_c =100$~fs. For the HEOM dynamics, we used $K=0$ and $L=5$ with $\delta t=1$ fs. Each site is coupled to a local bath where the coupling is described by a Debye spectral density. In addition to the vacuum level, the one and two exciton manifolds are included and the transition dipoles are taken as $+1.0$ and $-0.2$ to the two states, defining the $\mu_R$ and $\mu_L$ operators as the lower and upper diagonal matrices of
    \begin{equation}
     \mu =
        \begin{bmatrix}
            0 & 1.0 & -0.2 & 0 \\
            1.0 & 0 & 0 & -0.2 \\
            -0.2 & 0 & 0 & 1.0 \\
            0 & -0.2 & 1.0 & 0
        \end{bmatrix}.
    \end{equation}
    
    \item Excitation energy transfer dimer of Figs.~\ref{fig:damped} and~\ref{fig:extend_t2}: all parameters are the same with the exception of $\lambda = 50$~cm$^{-1}$ and $L=9$. 
\end{enumerate}

For completeness, the forms of the SB and EET models used here are as follows, 
\begin{equation}
    H = H_{\rm S} + H_{\rm B} + H_{\rm SB}.
\end{equation}
These terms correspond to the system, bath, and system-bath Hamiltonians. Both models have two sites, $N_\rmm{Hil}=2$. For the SB model, the system Hamiltonian has the form, 
\vspace{-10pt}
\begin{equation*}
    H_{\rm S} = \begin{bmatrix}
\varepsilon & \Delta \\
\Delta  & -\varepsilon
\end{bmatrix}.
\end{equation*}
Contrastingly, for the EET model up to the two-exciton manifold we use second quantization to write 
\begin{equation*}
H_{\rm S} = \sum_k \varepsilon_k B_k^\dagger B_k + \sum_{j,k} \Delta_{jk} B_j^\dagger B_k,
\end{equation*}
which can be written in the basis of many body exciton states: $\{\ket{00}, \ket{10}, \ket{01}, \ket{11} \}$:
\begin{equation*}
H_{\rm S} = \begin{bmatrix}
0 & 0 & 0 & 0 \\
0 & \varepsilon & \Delta & 0 \\
0 & \Delta & -\varepsilon & 0 \\
0 & 0 & 0 & 0 
\end{bmatrix},
\end{equation*}
where each site in $N_\rmm{Hil}$ can have $0$ or $1$ excitons. The commutation relations are $[B_j, B_k^\dagger]=\delta_{j,k}(1-2B_j^\dagger B_j)$.

For both models the bath is composed of independent harmonic oscillators. The main difference lies in the fact that for the SB model there is one antisymmetrically coupled bath, whereas in the EET model each site is connected to its local bath. Thus, for the SB model
\begin{equation}\label{eq:harm_bath}
    H_{\rm B} = \frac{1}{2}\sum_{n} [\hat{p}_n^2 + \omega_n \hat{q}_n^2],
\end{equation}
and
\begin{equation}
    H_{\rm SB} = \sigma_z \sum_{n} c_n \hat{q}_n,
\end{equation}
where $\sigma_z$ is the $z$ Pauli matrix, $\hat{q}_n$ and $\hat{p}_n$ are the mass-weighted position and momentum operators for the $n^\rmm{th}$ harmonic oscillator in the bath, $\omega_n$ is the frequency of the $n^\rmm{th}$ oscillator and $c_n$ its coupling constant to the spin. The couplings are given by the spectral density of the system, 
\vspace{-10pt}
\begin{equation}
    J(\omega) = \frac{\pi}{2}\sum_{n} \frac{c_n^2}{\omega_{n}}\delta(\omega - \omega_n).
\end{equation}

For the EET dimer model,
\begin{equation}
    H_{\rm B} = \sum_{k}^{{\rm N_{Hil}}} H_{\rm B}^{(k)},
\end{equation}
meaning $N_\rmm{Hil}$ independent baths with the form of Eq.~\ref{eq:harm_bath}. The on-site coupling is written as
\begin{equation}
    H_{\rm SB} = \sum_{k} B_k^\dagger B_k \sum_{n} c_{k,n} \hat{q}_{k,n},
\end{equation}
where now each harmonic oscillator contains two labels: the first, $k$, labels the electronic site to which it belongs, and the second, $n$, identifies that harmonic oscillator within the local bath. Similarly, the coupling constants of each site to its local bath, $c_{k,n}$, are given by the local spectral density,
\begin{equation}
    J_k(\omega) = \frac{\pi}{2}\sum_{n} \frac{c_{k,n}^2}{\omega_{k,n}}\delta(\omega - \omega_{k,n}),
\end{equation}
which are assumed to be the same for both sites. If the system contains two excitons (one on each site), these couple to both baths simultaneously.

\onecolumngrid
\vfill
\pagebreak
\twocolumngrid
\subsection*{References}
\vspace{-14pt}
\bibliography{Postdoc-2Dspec}

\end{document}